\documentclass{aa}
\usepackage{amsmath}

\usepackage[svgnames]{xcolor}
\definecolor{blue}{RGB}{0,0,255}
\usepackage{caption}
\usepackage[font={color=blue},figurename=Fig.,labelfont={it}]{caption}
\usepackage{bm}
\usepackage{subcaption}
\usepackage[varg]{txfonts}
\usepackage{graphicx}
\usepackage{natbib}
\usepackage{pdfpages}
\usepackage{wasysym}
\usepackage{rotating}
\bibpunct{(}{)}{;}{a}{}{,} 
\usepackage{hyperref}
\usepackage{indentfirst}
\usepackage{float}
\usepackage{lipsum} 
\usepackage{ulem}
\usepackage[symbol]{footmisc}
\hypersetup{
    colorlinks=true,                          
    linkcolor=blue, 
    citecolor=blue, 
    urlcolor=blue,
    linktoc=all
}

\newcommand{\eg}{\textit{eg. }}
\newcommand{\ie}{\textit{ie. }}
\newcommand{\cf}{\textit{cf. }}

\newcommand{\colorREF}[1]{{\color{blue}{#1}}}

\begin{document}
\title{On the Slow Drift of Solstices:  Milankovic Cycles and Mean Global Temperature}
\author{  F. Lopes\inst{1}
\and V. Courtillot\inst{1}
\and D. Gibert\inst{2}
\and J-L. Le Mouël \inst{1}
}
\institute{{Universit\'e Paris Cité, Institut de Physique du globe de Paris, CNRS UMR 7154, F-75005 Paris, France}
\and {LGL-TPE, Univ Lyon, Univ Lyon 1, ENSL, CNRS, UMR 5276, 69622, Villeurbanne, France}}

\date{}

\abstract {The Earth's revolution is modified by changes in inclination of its rotation axis. Despite the fact that the gravity field is central, the Earth's trajectory is not closed and the equinoxes drift. Changes in polar motion and revolution are coupled through the Liouville-Euler equations. \colorREF{Milankovic (1920)} argued that the shortest precession period of solstices is 20,700 years: the Summer solstice in one hemisphere takes place alternately every 11 kyr at perihelion and at aphelion. Milanković assumed that the planetary distances to the Sun and the solar ephemerids are constant. There are now observations that allow one to drop these assumptions. We have submitted the time series for the Earth’s pole of rotation, global mean surface temperature and ephemeris to iterative Singular Spectrum Analysis. iSSA extracts from each a trend, a 1yr and a 60yr component. Both the apparent drift of solstices of Earth around the Sun and the global mean temperature exhibit a strong 60yr oscillation. We monitor the precession of the Earth’s elliptical orbit using the positions of the solstices as a function of Sun-Earth distance. The “fixed dates” of solstices actually drift. Comparing the time evolution of the Winter and Summer solstices positions of the rotation pole and the first iSSA component (trend) of the temperature allows one to recognize some common features. A basic equation from Milankovic links the derivative of heat received at a given location on Earth to solar insolation, known functions of the location coordinates, solar declination and hour angle, with an inverse square dependence on the Sun-Earth distance. We have translated the drift of solstices as a function of distance to the Sun into the geometrical insolation theory of Milankovic. Shifting the inverse square of the 60yr iSSA drift of solstices by 15 years with respect to the first derivative of the 60yr iSSA trend of temperature, that is exactly a quadrature in time, puts the two curves in quasi-exact superimposition. The probability of a chance coincidence appears very low. Correlation does not imply causality when there is no accompanying model. Here Milankovic's equation can be considered as a model that is widely accepted. This paper identifies a case of agreement between observations and a mathematical formulation, a case in which an element of global surface temperature could be caused by changes in the Earth's rotation axis .}

\titlerunning{On the Slow Drift of Solstices}
\maketitle

\section{Introduction} 
For over a century and a half, geological evidence has been put forward to argue that Earth had undergone cyclical changes in climate \citep{Agassiz1837}. \cite{Adhemar1860} described the long periodicities associated with glacial and interglacial cycles that were explained sixty years later by the mathematical theory of climate due to \cite{Milankovic1920}.  A geometrical derivation shows that the insolation $W$ received at a location with coordinates $\varphi$ (latitude) and $\psi$ (longitude) is given by the (fundamental, yet simple) equation (eq. 20, page 15,  \cite{Milankovic1920}):
  
 \begin{equation}
     \dfrac{d W}{dt} = \dfrac{I_0}{\rho^2} [\sin \varphi \sin \delta + \cos \varphi \cos \delta \cos (\omega+\psi)]
     \label{Eq:01}
\end{equation}    
    where $\rho$ is the Sun-Earth distance, $\delta$ the Sun’s declination and $\omega$ its hour angle. It is generally considered that the shortest Milanković cycle is the 19kyr precession cycle (\eg \cite{Laskar2004, Lopes2021a}). The Earth’s eccentricity being quite small ($e =$ 0.016), it is legitimate at first to consider that, over such long durations $\rho$ and  $\delta$ are constant. Thus, \cite{Milankovic1920} assumes that all climate variations arise because of movements of the rotation axis (\ie of the pole).\\

    To first order, the rotation axis is perturbed by the conjugate effects of the Moon and Sun (\eg \cite{Laplace1799,Poincare1899}), resulting for instance in the luni-solar tides (\textit{eg.} \cite{Ray2014}). Over longer periods, the Jovian planets are the main source of perturbations of the rotation axis (\cf \cite{Laskar1993, Laskar2004}), giving rise to variations in precession, obliquity and eccentricity. \\
    
         It has recently been shown, based on observations, that the Jovian planets exert perturbations both on Earth and in the Sun at much shorter periods (centuries to years and less) (\eg \cite{Morth1979, Courtillot2021, Lopes2021b, Bank2022, Scafetta2022, Yndestad2022}). \\
     
Another source of perturbation of the Earth’s rotation axis is the precession of equinoxes, associated with Kepler’s second law (conservation of areas). There are four sources of precession of Earth’s equinoxes, that is rotation of the whole orbit (revolution).\\
     
1) The first is associated with Kepler’s laws. In the case of a central field and an elliptical orbit, for the orbit to be closed it is necessary and sufficient that the orbit’s angular change after $n$  revolutions be of the form $\Delta \varphi = 2\pi m/n$, where $m$ is the number of full revolutions necessary for the planet to recover its initial position. There are only two central fields in which $\Delta \varphi$ is a rational fraction of $2\pi$, ensuring closed orbits, that is fields in $r^2$ and $1/r$ the latter being the case of our solar system (\cf \cite{Landau1964}).\\

2) The second involves the joint effects of the Moon and Sun. Let us quote \colorREF{d’Alembert} (\colorREF{1749}, page 14): “\textit{Enfin, l’inclinaison de l’axe terrestre au plan de l’ecliptique doit modifier aussi l’action du Soleil; car selon que cet axe sera différemment incliné, il fera à chaque point de l’ecliptique un angle différent avec la ligne qui joint les centres de la Terre et du Soleil; par conséquent la quantité et la loi de l’action du Soleil, dépend de l’inclinaison de l’axe, et c’est aussi ce que l’analyse apprend}.” \\

    As clearly stated by \cite{Laplace1799} and later \cite{Poincare1899}, the planet’s revolution is modified by changes in inclination of the rotation axis, principally due to the joint actions of the Moon (for 2/3rds) and Sun (for 1/3rd). One is no longer in case 1: despite the fact that the field is a central one, the trajectory is not closed anymore. \colorREF{d’Alembert} (\colorREF{1749}, page 52) estimates the drift of the equinoxes to be about 50 ’’ per year, that is a precession period of 25920 yr. \\

    The two other processes are relativistic. 
    
    3) The Sun containing 99\% of the total mass of the solar system, \cite{Schwarzschild1916} shows that the planet’s revolution about the Sun produces an additional precession of about 3.8" per century, or a period of some 33 million years. \\
    
    4) Because the Sun is actually a huge rotating mass, there is an additional relativistic component of precession, with a period on the order of 5.8 million years \cite{Lense1918}. \\
  
        For Newton, planetary bodies attract (or repel) the oceans (and atmospheres) and this re-organization of masses modifies the rotation axis. For d’Alembert, Lagrange, Laplace and Poincaré, Changes in polar motion and revolution are coupled and involve the luni-solar torques, as in a top. Re-organization of Earth’s fluid envelopes (\eg tides) follows.\\
        
    As \cite{Milankovic1920} writes, precession of the equinoxes is actually due to the joint attraction of the Moon and Sun on the Earth’s equatorial bulge and its period is in theory 26,000 yr. Because the Earth itself rotates, areolar velocities vary between perihelion and aphelion (Kepler’s second law); because of centrifugal forces, a precession with a period of 19,000 yr appears. So the precession of equinoxes undergoes a double periodicity, with a mean of 22,500 yr (half period 11,250 yr). Indeed, \colorREF{Milankovic} (\colorREF{1920}, p.221) writes that the first precession of perihelion (for us solstices) is 20,700 Julian years and that the consequence of this precession is that the Summer solstice in one hemisphere (when that hemisphere receives maximum insolation) takes place alternately every 11,000 yr at perihelion (thus a warmer Summer) and aphelion (thus a cooler Summer). The difference in insolation (energy received by Earth) between maximum and minimum is a function of eccentricity. \\

    The 26 kyr period of precession has first been determined in the frame of Newtonian physics by \cite{dAlembert1749}. It is rather close to the first precession cycle of 19 kyr in \colorREF{Milanković’s (1920)} theory. When Milanković makes the assumption that the planetary distances to the Sun and the solar ephemerids are constant, he can estimate climate maxima but not their smooth transitions between equinoxes and solstices. Today, we have access to observations that allow one to drop the hypotheses that $\rho$, $\delta$ and $\omega$ in equation \ref{Eq:01} are constant. Thus, we can evaluate the consequences of changes in the position of the rotation axis on, for instance, atmospheric temperature, that is the main parameter in \colorREF{Milanković’s (1920)} theory of climate.

\section{The data: temperature, pole motion, and solar ephemerids}
\begin{figure}[htb]
    \centering
	\begin{subfigure}[b]{\columnwidth}
		\centerline{\includegraphics[width=\columnwidth]{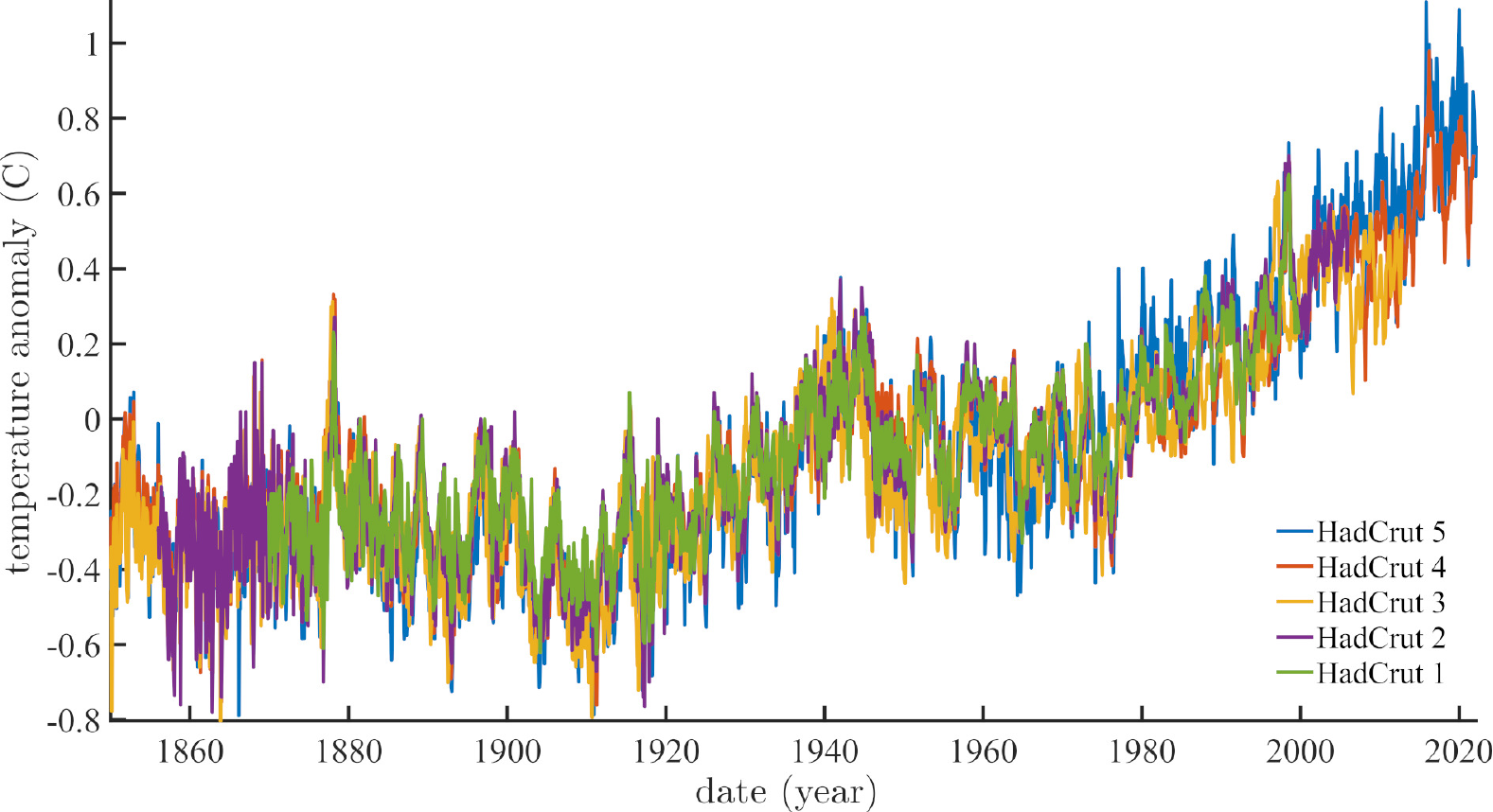}} 	
		\subcaption{The five mean global temperature data sets \textbf{HadCrut1} to \textbf{HadCrut5} from 1850 to the present maintained by the \textit{Hadley Center for Climate Prediction and Research} (see text).}
		\label{Fig:01a}
	\end{subfigure}
	\begin{subfigure}[b]{\columnwidth}
		\centerline{\includegraphics[width=\columnwidth]{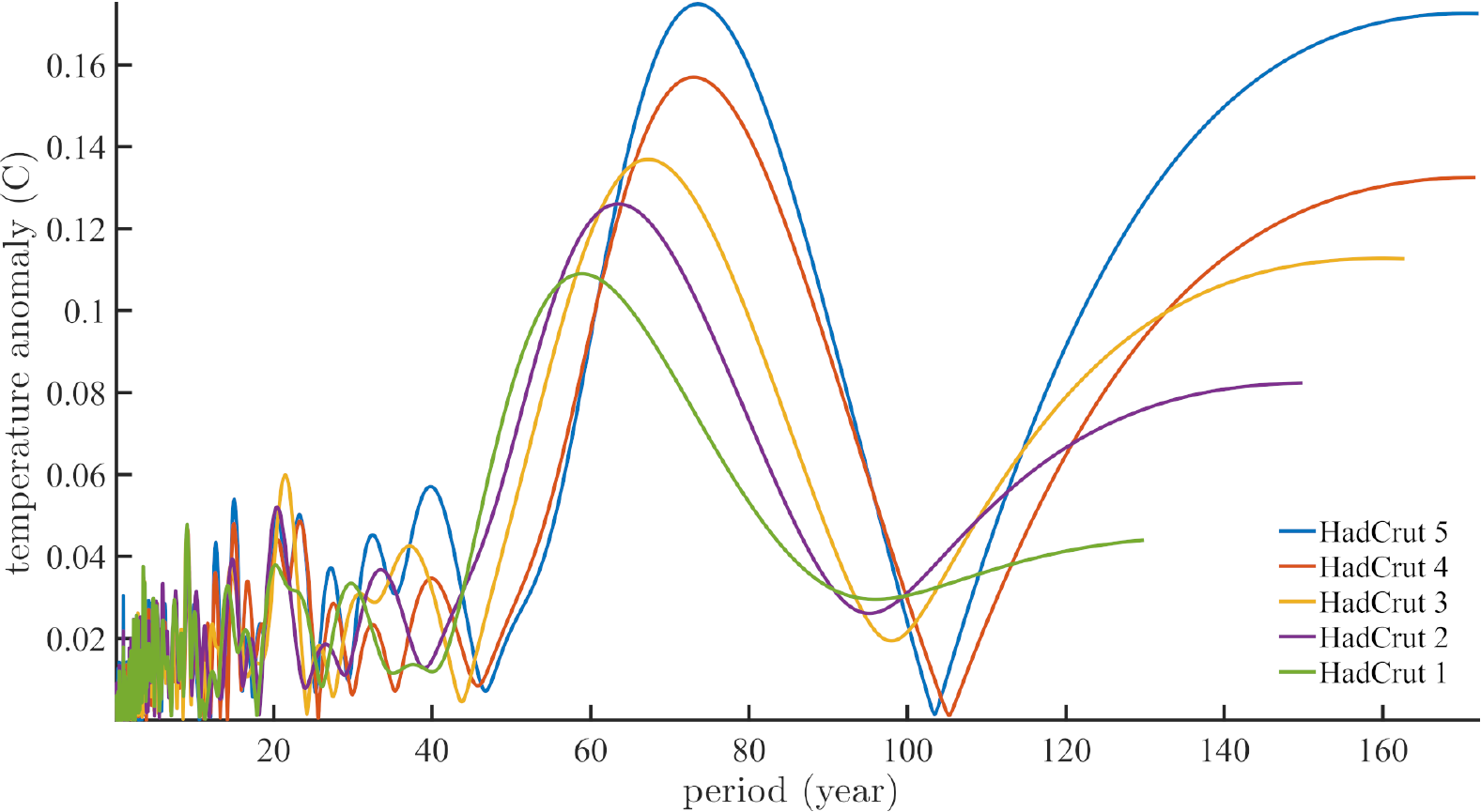}} 	
		\subcaption{The 5 Fourier spectra of the 5 data sets in Figure \ref{Fig:01a}}
		\label{Fig:01b}
	\end{subfigure}
	\caption{The five \textbf{HadCrut} mean global temperature since 1850}
	 \label{Fig:01}
\end{figure}

\textit{2-1: Mean global temperatures: }We have used the data series maintained by the \textit{Hadley Center for Climate Prediction and Research} under the name HadCrut. In order to have an idea of the reliability of the data, we have selected five successive sets of \textbf{HadCrut} data: HadCrutv\footnote{https://crudata.uea.ac.uk/cru/data/crutem1/}, 1870-2000 \cite{Rayner1996}; HadCrut2\footnote{https://crudata.uea.ac.uk/cru/data/crutem2/}, 1856-2006 \cite{Rayner2003}; HadCrut3\footnote{https://crudata.uea.ac.uk/cru/data/crutem3/}, 1850-2014 \cite{Brohan2006}; HadCrut4\footnote{https://crudata.uea.ac.uk/cru/data/crutem4/}, 1850-2021 \cite{Osborn2014} and HadCrut5\footnote{https://crudata.uea.ac.uk/cru/data/temperature/} 1850-2022 \cite{Osborn2021}. Figure \ref{Fig:01a} shows all the data, and Figure \ref{Fig:01b} their Fourier transforms. There are rather significant differences between the data series, for instance between 1940 and 2020 in HadCrut3 (yellow curve) vs HadCrut5 (blue curve). Differences become larger after 1950, to the point that HadCrut3 has a plateau after 2000 when HadCrut5 grows linearly since 1960. We have already worked on these data sets \cite{LeMouel2020a} and pointed out these differences \cite{Courtillot2013}, Figure 4. These differences of course result in differences in the Fourier spectra of Figure \ref{Fig:01}. As a result, the dominant spectral peak shifts from 60 to 80 yr, a topic discussed in several papers \citep{Mazzarella2012, Courtillot2013, Gervais2016, Veretenenko2019, Scafetta2020}. 

\textit{2-2: Solar ephemerids:} We have obtained the Sun’s ephemerids from 1846 to the present from Institut de Mécanique Céleste et du Calcul des Ephémérides (\textbf{IMCCE}\footnote{http://vo.imcce.fr/webservices/miriade/?forms}). We do not present a figure with the raw data: the Earth orbit’s eccentricity is so small that an annual oscillation since 1846 would transform into an unreadable quasi-sinus with 176 oscillations on the 15cm (or so) width of the figure, that is 14 oscillations per cm. \\

\begin{figure}[htb]
    \centering
	\begin{subfigure}[b]{\columnwidth}
		\centerline{\includegraphics[width=\columnwidth]{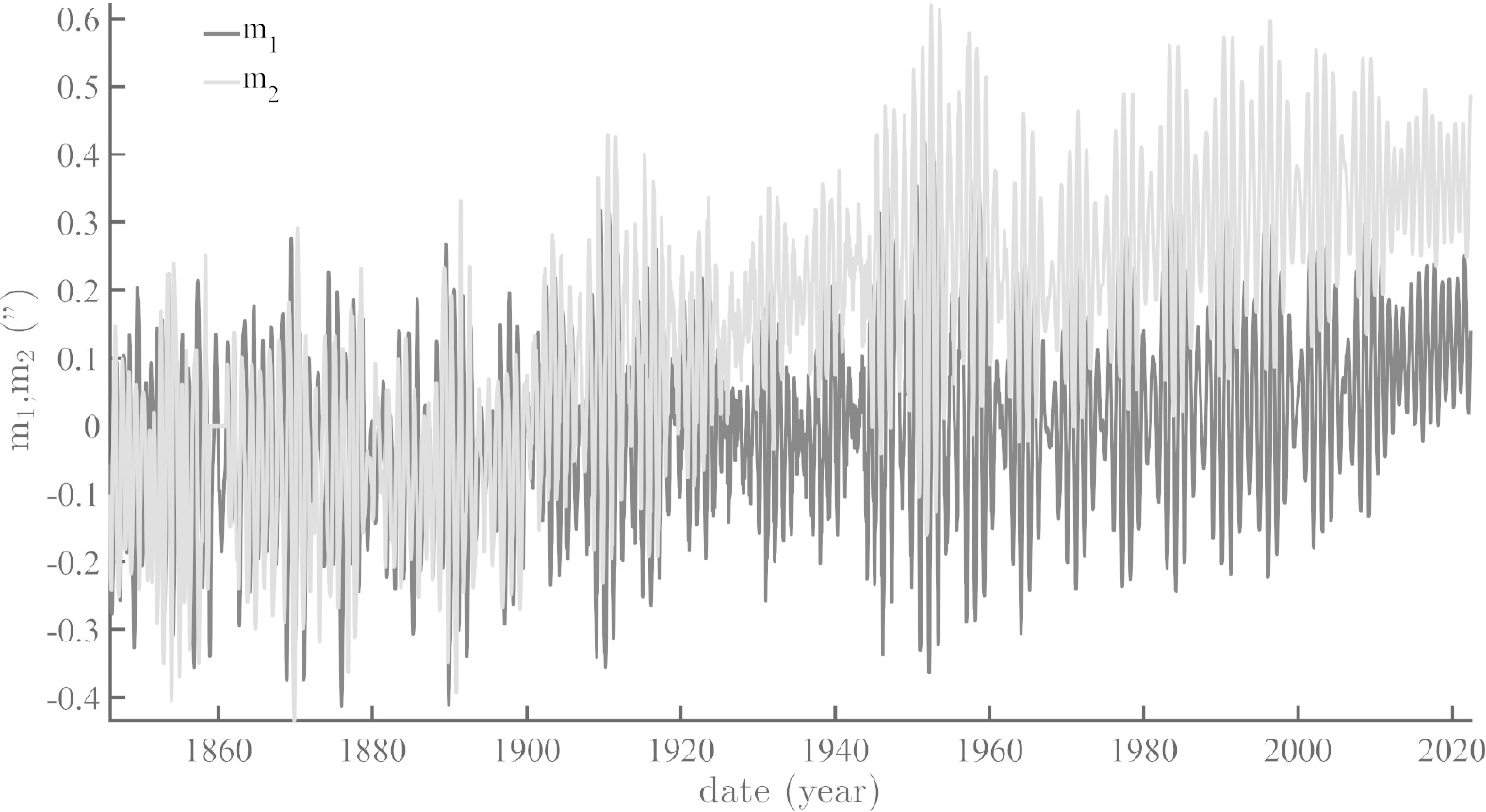}} 	
		\subcaption{Evolution of the couple ($m_1$, $m_2$) since 1846 (\textbf{IERS} data).}
		\label{Fig:02a}
	\end{subfigure}
	\begin{subfigure}[b]{\columnwidth}
		\centerline{\includegraphics[width=\columnwidth]{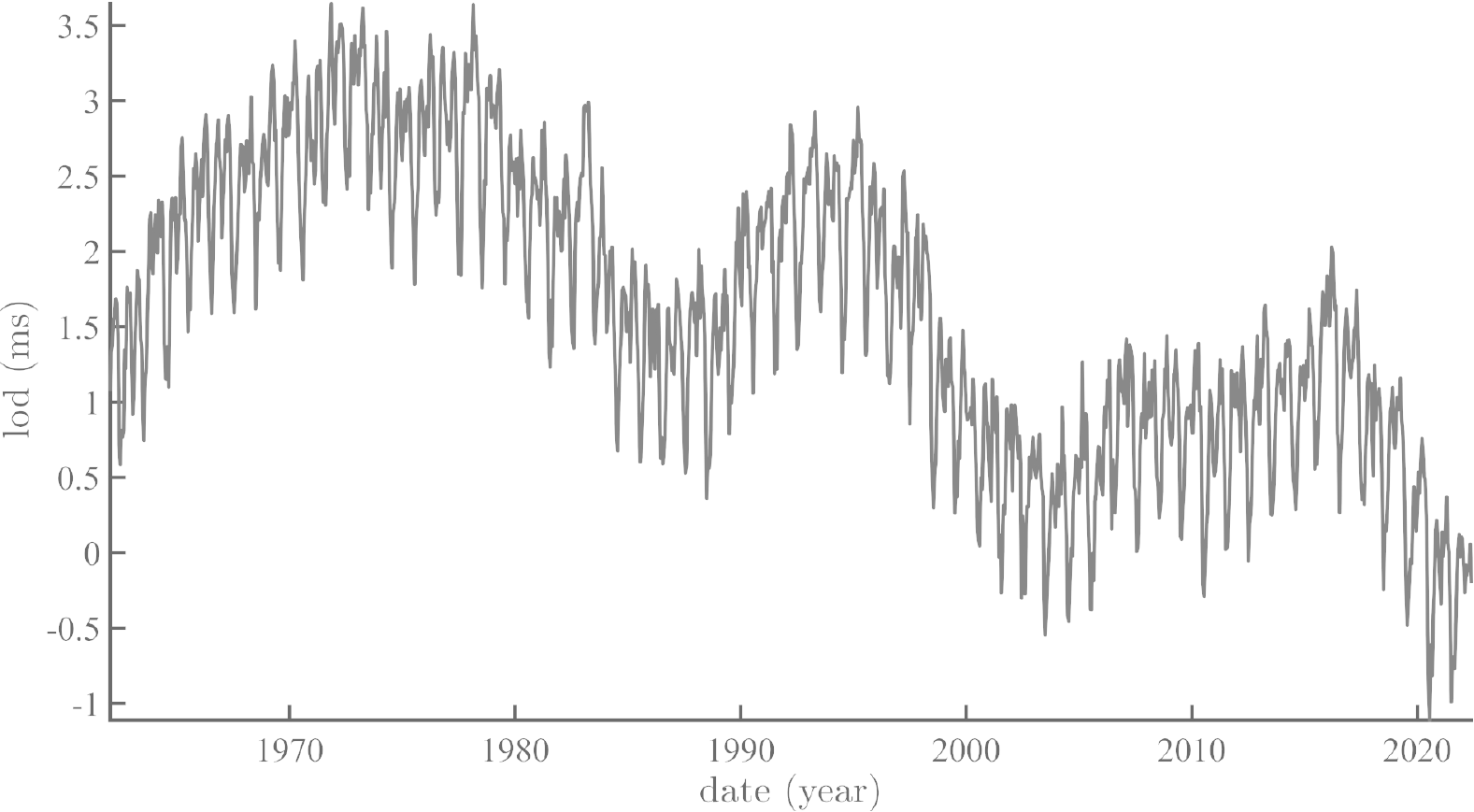}} 	
		\subcaption{Evolution of lod since 1962 (\textbf{IERS} data).}
		\label{Fig:02b}
	\end{subfigure}
	\begin{subfigure}[b]{\columnwidth}
		\centerline{\includegraphics[width=\columnwidth]{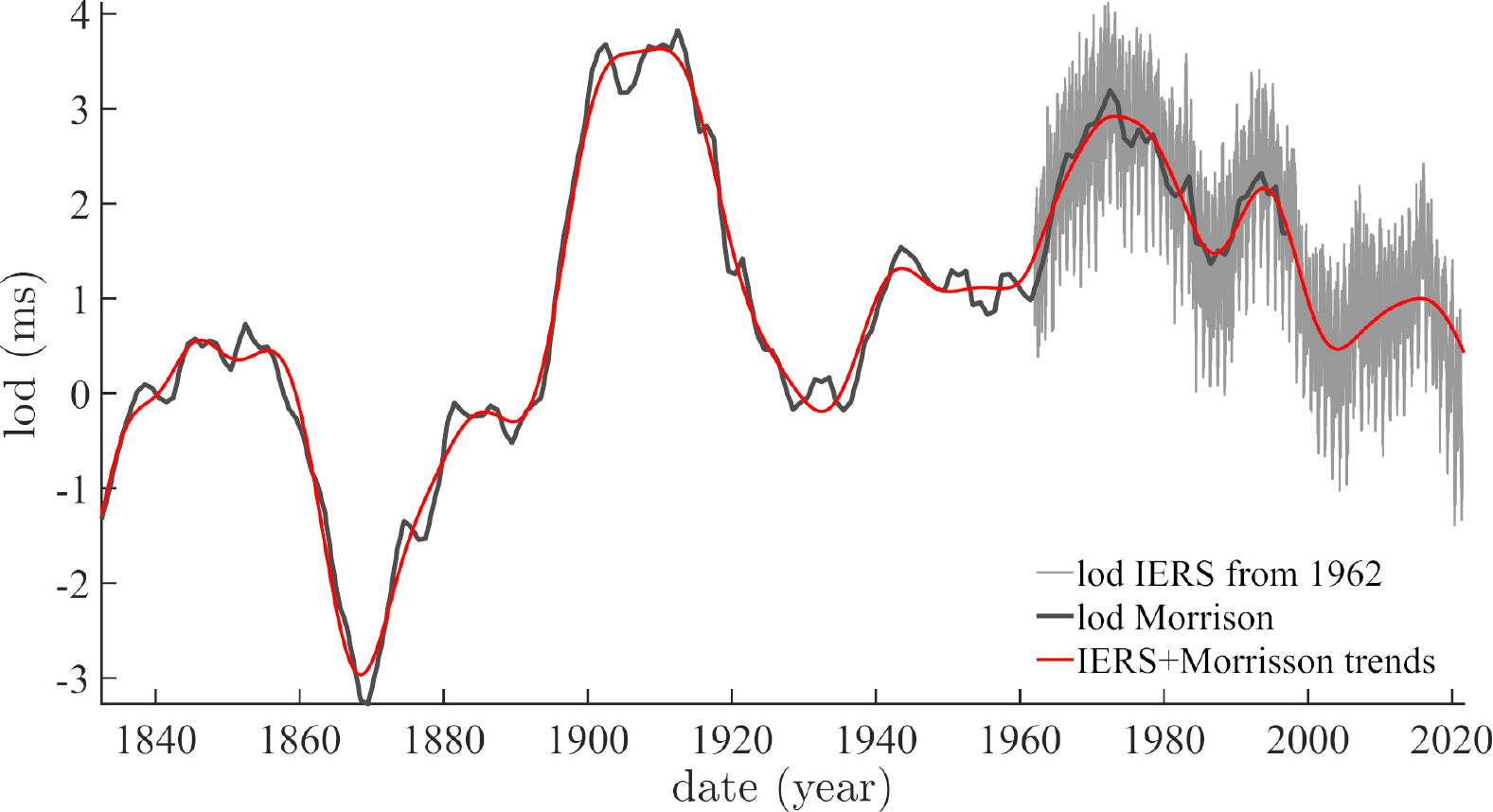}} 	
		\subcaption{Black curve: lod semi-annual data since 1832 (from \cite{Stephenson1984}); Gray: daily lod data since 1962 from \textbf{IERS}; Red curve: The median of the trend for these two combined data sets.}
        \label{Fig:02c}
	\end{subfigure}
	\caption{Pole Motion and length of day from IERS data.}
	 \label{Fig:02}
\end{figure}

\textit{2-3: Rotation pole and length of day:} The motions of the rotation pole and variations in rotation velocity are available at the \textit{International Earth Rotation Reference Systems Service}\footnote{ https://www.iers.org/IERS/EN/DataProducts/EarthOrientationData/eop.html}(\textbf{IERS}). They consist in the couple of coordinates ($m_1$, $m_2$) of motion of the rotation pole \textbf{PM} (see \cite{Lambeck2005, Lopes2021b}) and the length of day \textbf{lod} (\eg  \cite{LeMouel2019a}). We have selected data set EOP C01 IAU19801. Figures \ref{Fig:02a} and \ref{Fig:02b} respectively show the evolution of the couple ($m_1$, $m_2$) since 1946 and of \textbf{lod} since 1962. We have used the semi-annual \textbf{lod} data provided by \cite{Stephenson1984} for the period 1832-1997, combined with the \textbf{IERS} data, resulting in the mean curve between 1832 and 2022 shown in Figure \ref{Fig:02c} \citep{Lopes2022}.

\section{Extraction and Analysis of the trends and annual oscillations}
As in a number of previous papers \eg \cite{LeMouel2019b,LeMouel2020b,Lopes2021b}, we have submitted time series to iterative Singular Spectrum Analysis (\textbf{iSSA}) and we will now do the same for the rotation, temperature and ephemeris time series presented in the previous section. We refer the reader to these papers and to \cite{Golyandina2013} for the \textbf{SSA} method, to \cite{Lemmerling2001} for properties of the Hankel and Toeplitz matrices that it uses, and to \cite{Golub1971} for the singular value decomposition algorithm \textbf{SVD}).\\

\begin{figure}[htb]
    \centering
	\begin{subfigure}[b]{\columnwidth}
		\centerline{\includegraphics[width=\columnwidth]{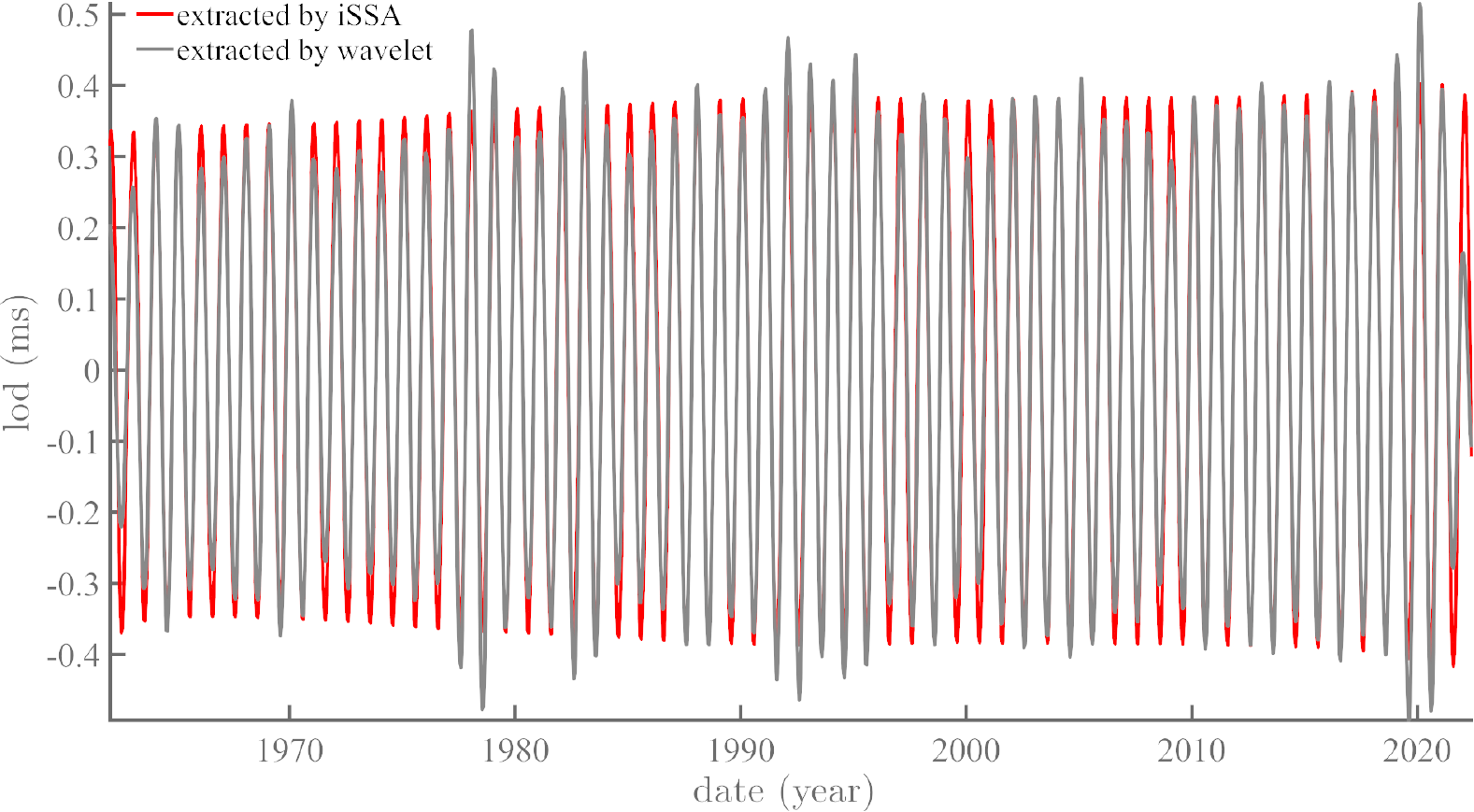}} 	
		\subcaption{Black curve: annual component of \textbf{lod} extracted by the method of continuous wavelets; Red curve: annual component of \textbf{lod} extracted by the \textbf{iSSA} method.}
		\label{Fig:03a}
	\end{subfigure}
	\begin{subfigure}[b]{\columnwidth}
		\centerline{\includegraphics[width=\columnwidth]{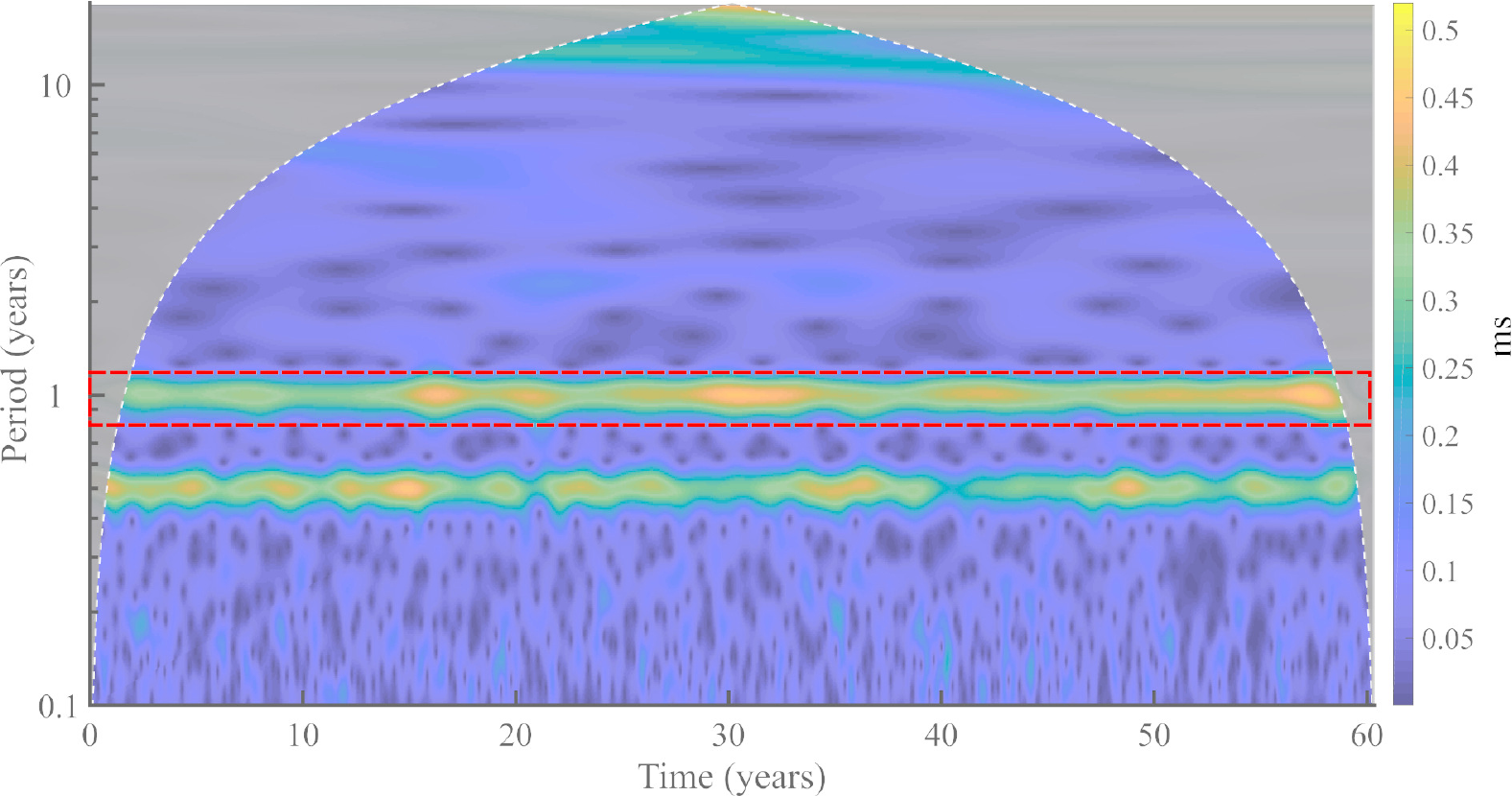}} 	
		\subcaption{Scalogram of \textbf{lod} since 1962. The wavelet transform is not defined in the gray zone. The red dashed lines enclose the wavelet coefficients corresponding to the 1 yr period.}
		\label{Fig:03b}
	\end{subfigure}
	\caption{Extraction of oscillatory component by continues wavelet transform: an example.}
	 \label{Fig:03}
\end{figure}

    Figure \ref{Fig:03a} shows the first oscillatory \textbf{iSSA} component of lod that has a period of 1 yr. In order to check the quality of \textbf{iSSA}, we have compared some of the \textbf{iSSA} results with those obtained with the method of continuous wavelets, that is widely used in the literature. Figure \ref{Fig:03b} shows the scalogram (\ie the continuous wavelet transform) of \textbf{lod} since 1962 (Figure \ref{Fig:02b}). We have selected a Morse wavelet (\cf \cite{Olhede2002, Lilly2012}), an analytic wavelet that is well adapted to time series with variable spectra. Between the dashed red lines are the ordinates of the wavelet coefficients associated with the 1 yr oscillation. Given the property of information redundancy of the wavelet kernels, one can extract the ridge in the scales covered by the 1 yr periodicity and reconstruct the signal \cite{Gibert1998}. The (wavelet) reconstructed annual component of lod is shown as the black curve in Figure 03a. The comparison is good but not perfect: the modulation of the iSSA curve (in red) is smoother than that of the wavelet reconstruction. The similarity of the two curves in Figure \ref{Fig:03a} together  give us confidence that \textbf{iSSA} can correctly extract the annual components of the three time series introduced in Section 2. \\

    A difficulty seen clearly in Figure \ref{Fig:03b} is that the signal’s energy can spread and diffuse over several scales. The reconstruction could be optimized by applying correction methods, such as the reassignment method (\cf  \cite{Auger1995}), but this would require an additional step that is not necessary in the present study.

\section{The Lissajous diagrams}
    The main points in space that can be used to monitor the precession of an elliptical orbit are the solstices and equinoxes. The fixed dates of their occurrences are December 21 for the Winter solstice, March 21 for the Spring equinox, June 21 for the Summer solstice, and September 23 for the Fall equinox. Since the legal and astronomical calendars are not exactly the same, this entails an error on their positions that can be estimated. The variations of these positions are actually very small: the perimeter of the Earth’s orbit being close to 6.28 astronomical units (a.u.), the December 21 positions are at a distance of 0.98412 $\pm$ 3.46 10$^{-5}$ a.u., that is an error of 5.5 10$^{-4}$ between 1844 and 2022. The error is 5.8 10-4 for the other solstice and close to 7.9 10$^{-4}$  and 8.7 10$^{-4}$  for the two equinoxes. \\
\begin{figure}[htb]
    \centering
	\begin{subfigure}[b]{\columnwidth}
		\centerline{\includegraphics[width=\columnwidth]{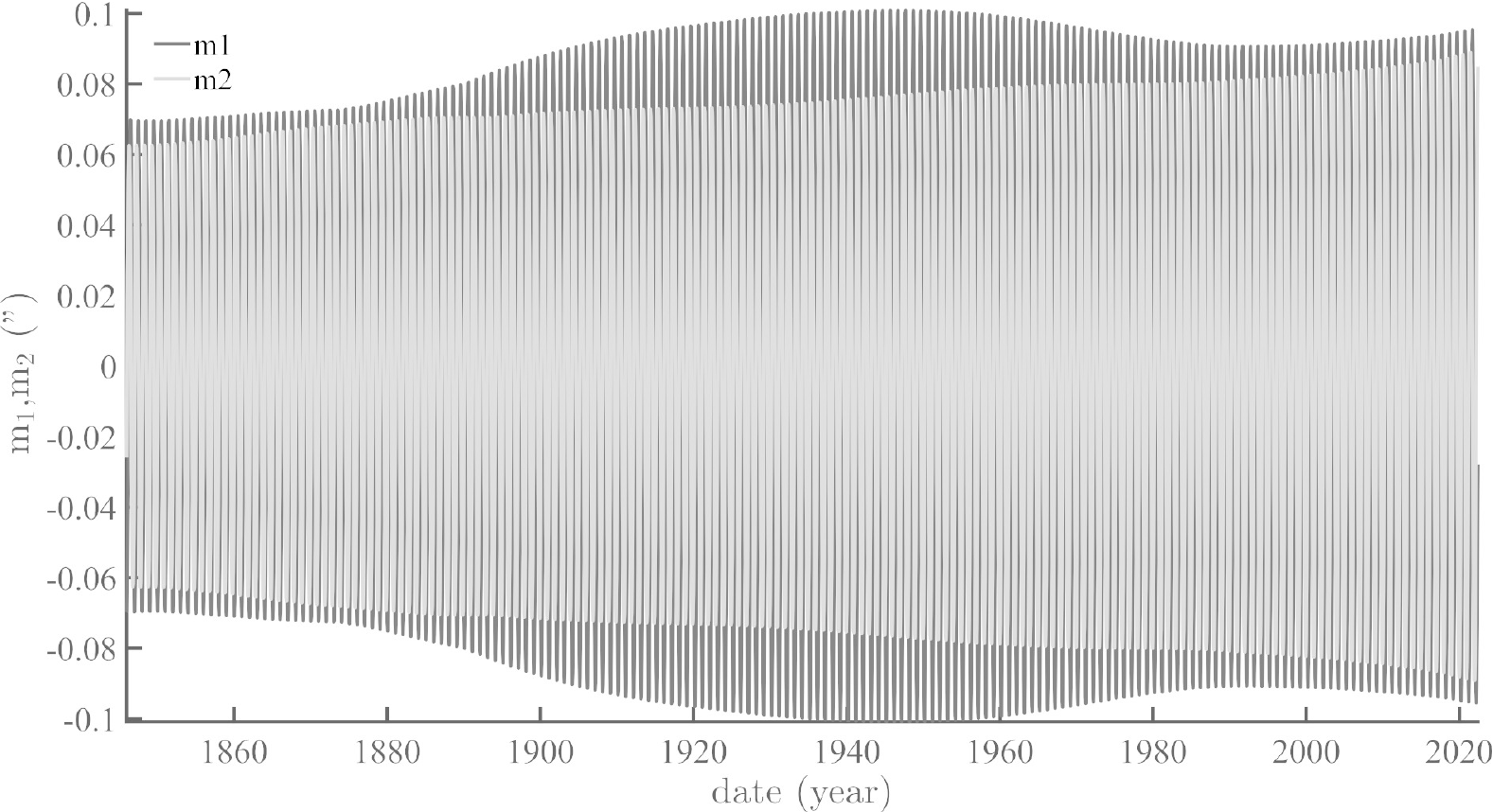}} 	
		\subcaption{The annual couple ($m_1$ , $m_2$) of polar motion coordinates extracted by \textbf{iSSA}.}
		\label{Fig:04a}
	\end{subfigure}
	\begin{subfigure}[b]{\columnwidth}
		\centerline{\includegraphics[width=\columnwidth]{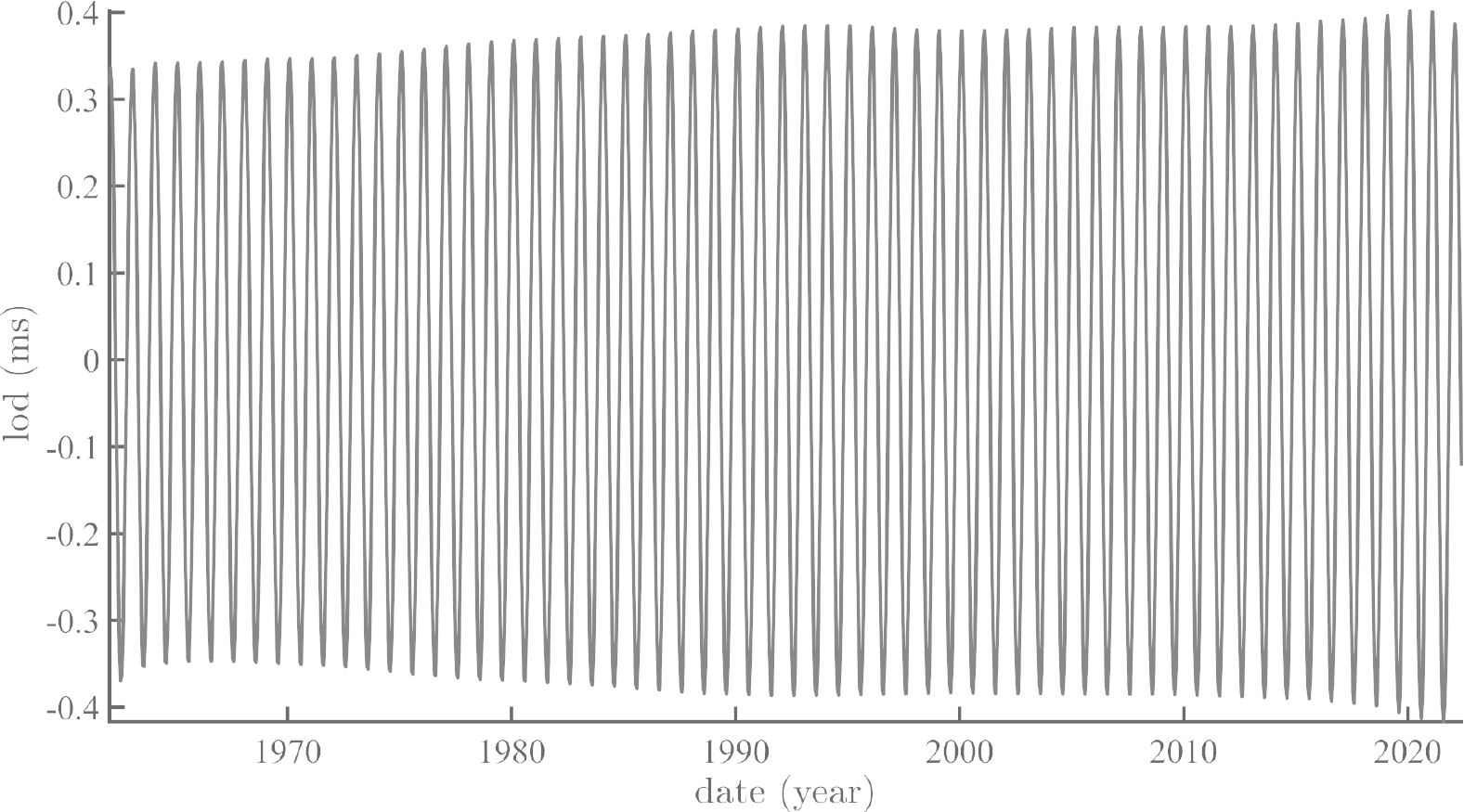}} 	
		\subcaption{The annual component of \textbf{lod} extracted by \textbf{iSSA}.}
		\label{Fig:04b}
	\end{subfigure}
	\caption{Annual components extractred from pole path and \textbf{lod}.}
	 \label{Fig:04}
\end{figure}   
    
\begin{figure}[H]
    \centering
	\begin{subfigure}[b]{\columnwidth}
		\centerline{\includegraphics[width=8cm]{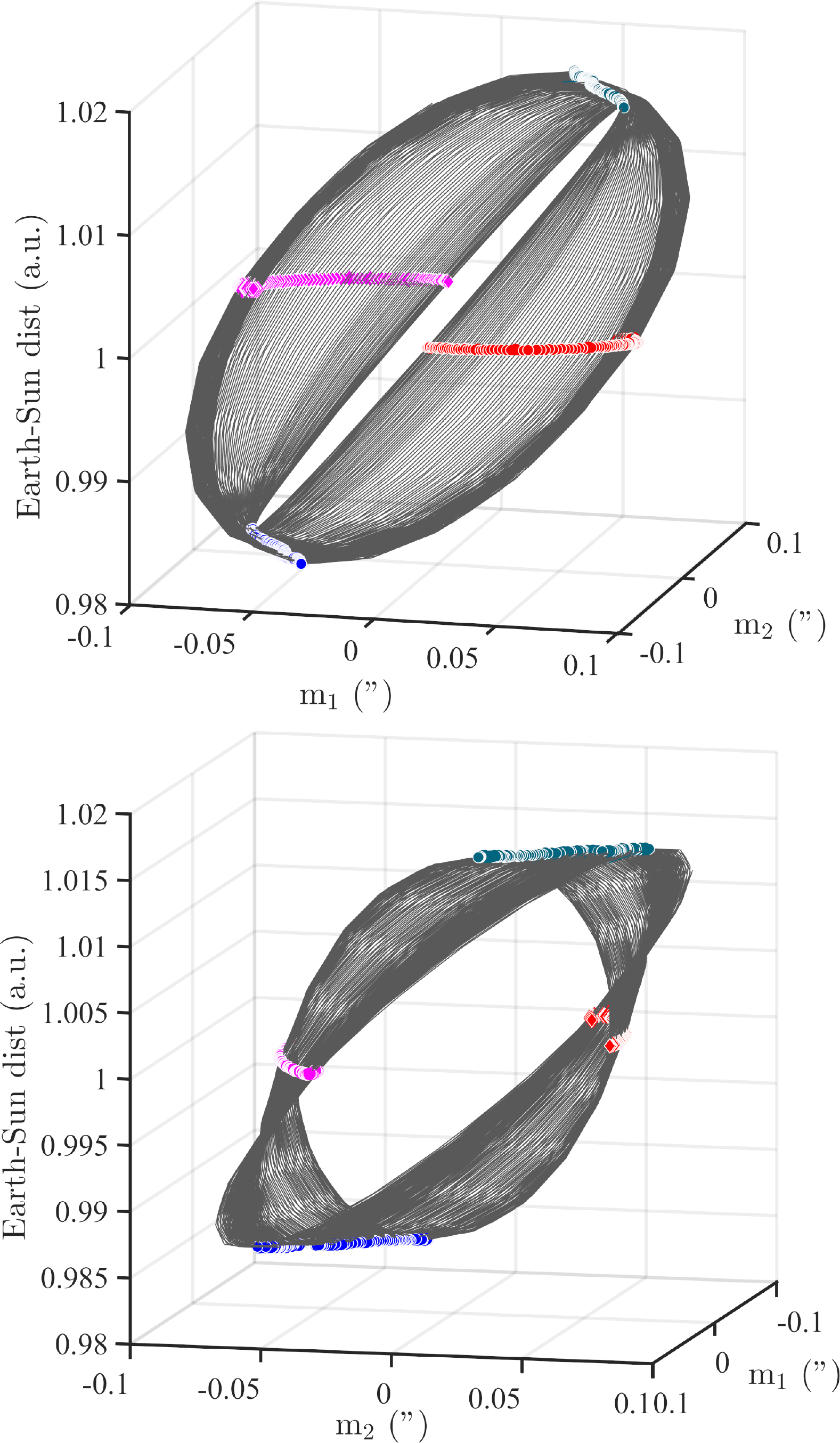}} 	
		\subcaption{Lissajous diagram for the couple of rotation pole motion coordinates ($m_1$ , $m_2$) as they vary with Sun-Earth distance, i.e. as a function of time.}
		\label{Fig:05a}
	\end{subfigure}
	\begin{subfigure}[b]{\columnwidth}
		\centerline{\includegraphics[width=8cm]{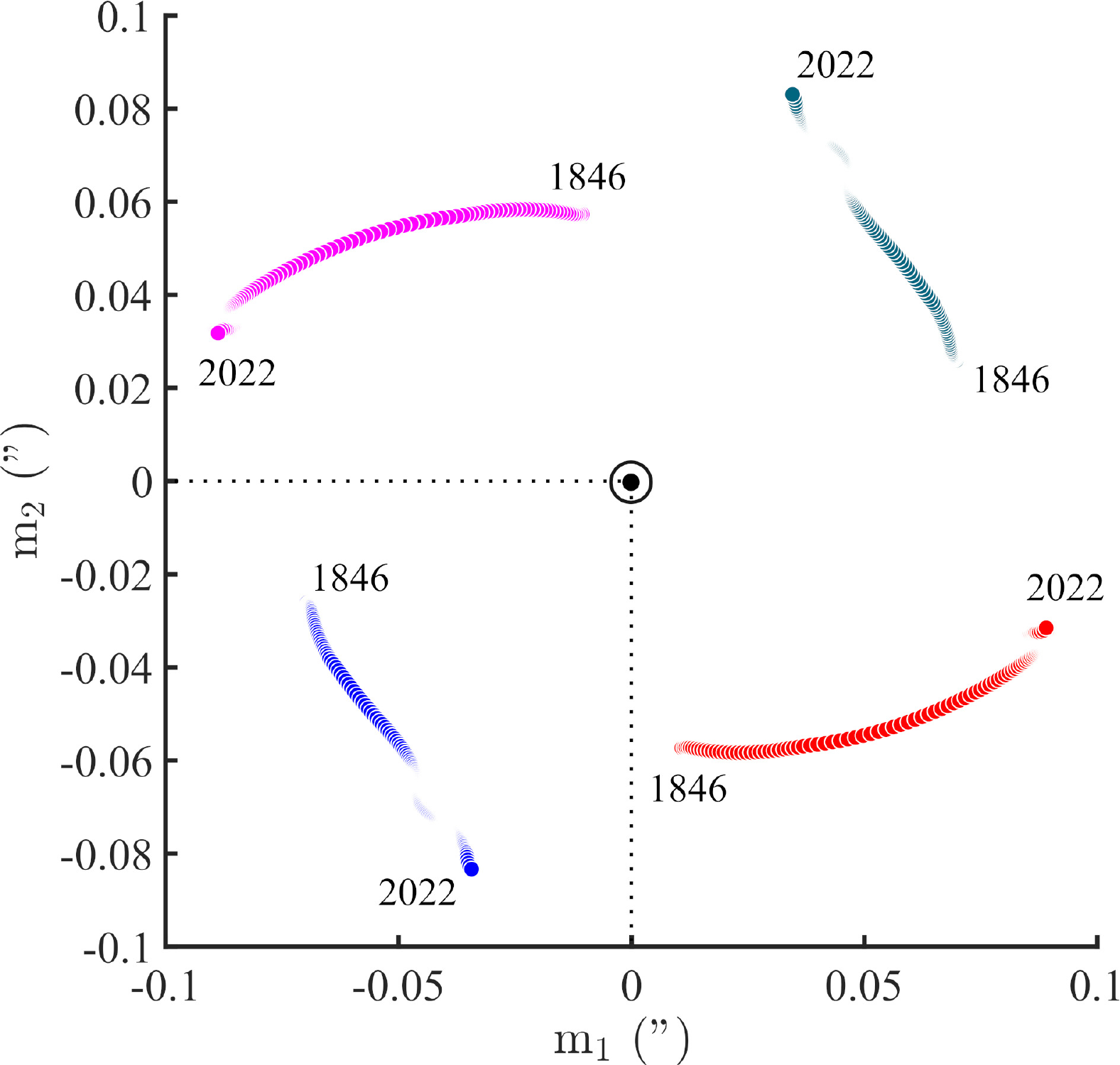}} 	
		\subcaption{Projection of the drift of equinoxes and solstices from 1846 to the present in the ($m_1$ , $m_2$) plane.}
		\label{Fig:05b}
	\end{subfigure}
	\caption{Drift of equinoxes and solstices since 1846.}
	 \label{Fig:05}
\end{figure}

Figure  \ref{Fig:05a} shows the evolution of the trajectories of Figure 4a as a function of Sun-Earth distance (ephemerids), that is in other words as a function of time. We call these by analogy to Lissajous orbits in astronomy “Lissajous diagrams”. The Lissajous diagram of ($m_1$, $m_2$) is shown in two perspectives in order to gain some insight on its topology. The locations of the four remarkable points (equinoxes and solstices) are shown in four different colors (the same are used throughout of the paper). We can see that the closer the Earth is to the Sun, the more the rotation axis straightens; the farther it is, the larger the amplitudes of motions and the flatter the rotation axis, i.e. the larger its declination (see also Figure  \ref{Fig:05b}). Note that the “fixed dates” of equinoxes and solstices appear to drift as a function of time. 

The conservation of momentum of the orbiting planet implies that its areolar velocity is constant (the areolar velocity is the area spanned by the vector radius – the Sun to Earth vector - per unit time). As a consequence of this “law of areas”, the orbital velocity varies from a maximum of 30.29 km/s at perihelion to a minimum of 29.29 km/s at aphelion. As explicitly stated by \cite{Lagrange1781, Lagrange1782, Laplace1799} and \cite{Poincare1899} long ago, and recently re-emphasized by \cite{Lopes2021a, Courtillot2022} and \cite{Lopes2022}, the iSSA annual component of polar motion ($m_1$, $m_2$) is controlled by variations in Sun-Earth distance $d_{SE}$. \\

\begin{figure}[H]
    \centering
	\begin{subfigure}[b]{\columnwidth}
		\centerline{\includegraphics[width=7cm]{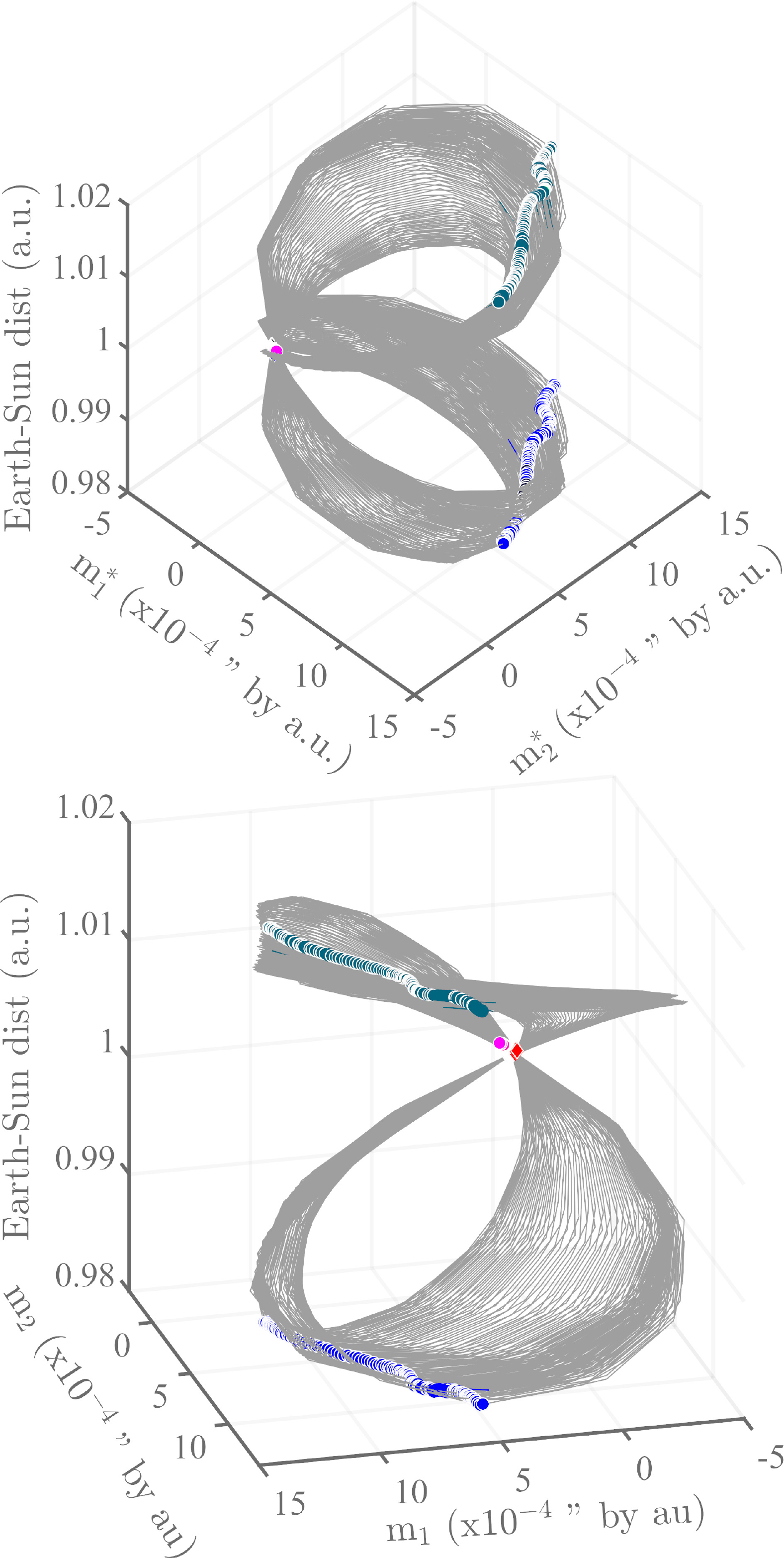}} 	
		\subcaption{Lissajous diagram for the couple of coordinates ($m_1^*$, $m_2^*$) as they vary with Sun-Earth distance, \ie as a function of time.}
		\label{Fig:06a}
	\end{subfigure}
	\begin{subfigure}[b]{\columnwidth}
		\centerline{\includegraphics[width=7cm]{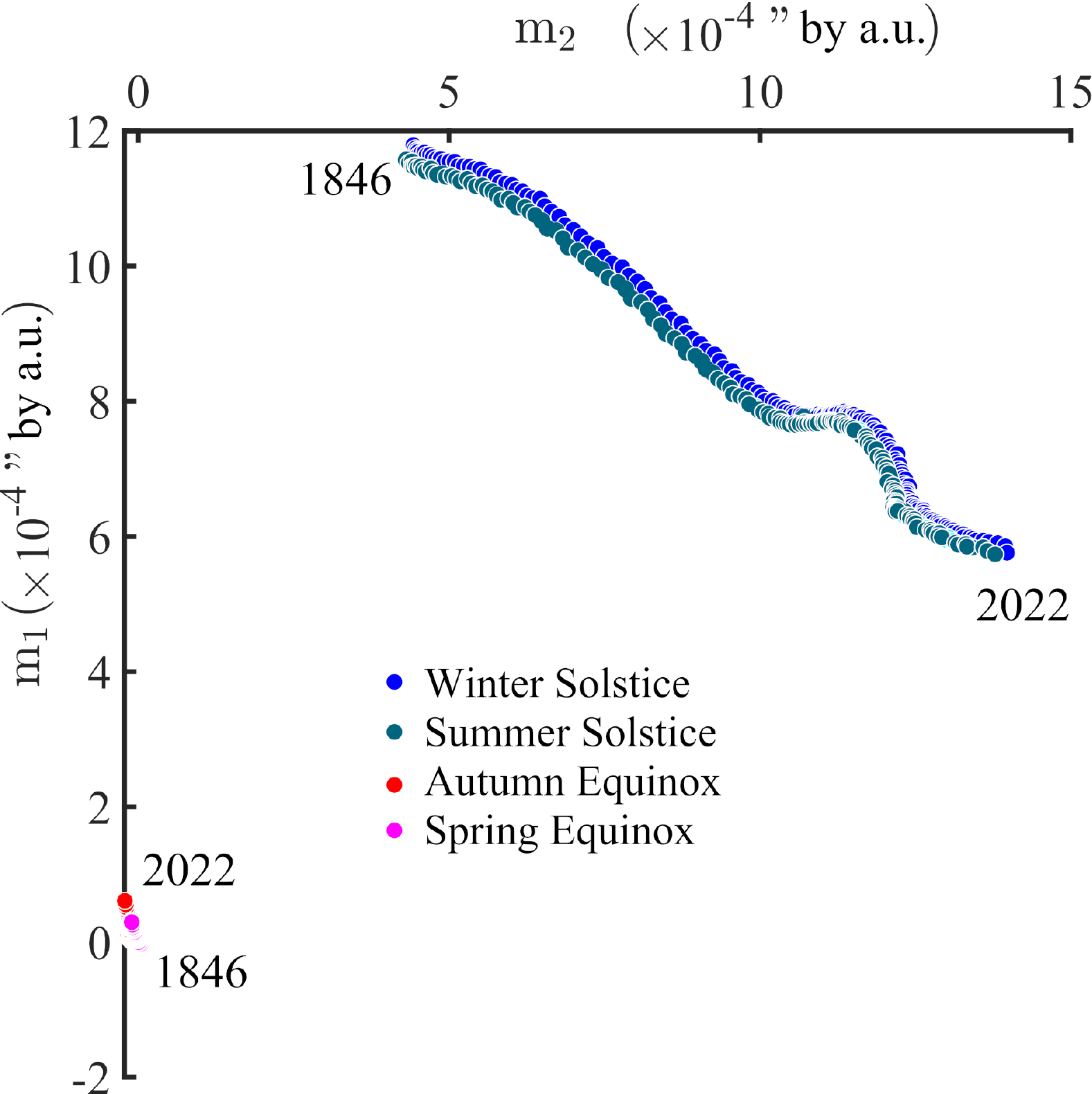}} 	
		\subcaption{Projection of the drift of equinoxes and solstices from 1846 to the present in the ($m_1^*$, $m_2^*$) plane.}
		\label{Fig:06b}
	\end{subfigure}
	\caption{Drift of equinoxes and solstices since 1846 with the new parameters $m_1^*$ and $m_2^*$}
	 \label{Fig:06}
\end{figure}

    In order to emphasize the relative amplitudes of the drift and the butterfly-like shape of the diagram, we have actually multiplied ($m_1$, $m_2$) by the centered value $d_{SE}^* = d_{SE} – mean(d_{SE})$. This also makes it clear that the drift of solstices is larger than that of equinoxes. Figure \ref{Fig:06a} shows the Lissajous diagram equivalent to that in Figure 05a for the couple ($m_1^*$,$m_2^*$). The Lissajous diagrams of Figure \ref{Fig:06a} represent the geodesic evolution of Kepler areas. \\
    
    One sees clearly in Figure  \ref{Fig:06a} that polar motion reaches a minimum at the equinoxes (red and orange dots) when solar attraction is the weakest, and a maximum at the solstices (green and blue dots) when the Sun, Earth and focus of the ellipse are aligned. We shall see below that the same applies to \textbf{lod}.\\
    
    These results  actually require another small correction, in relation with Kepler’s second law. The conservation of momentum of the orbiting planet implies that its areolar velocity is constant (the areolar velocity is the area spanned by the vector radius – the Sun to Earth vector - per unit time). As a consequence of this “law of areas”, the orbital velocity varies from a maximum of 30.29 km/s at perihelion to a minimum of 29.29 km/s at aphelion. We introduce new more physical variables by multiplying the polar motion coordinates by the Sun-Earth distance $d_{SE}$:
\begin{equation*}
m_1^* =m_1*d_{SE}, \quad m_2^* =m_2*d_{SE}
\end{equation*}

\begin{figure}[H]
    \centering
	\begin{subfigure}[b]{\columnwidth}
		\centerline{\includegraphics[width=\columnwidth]{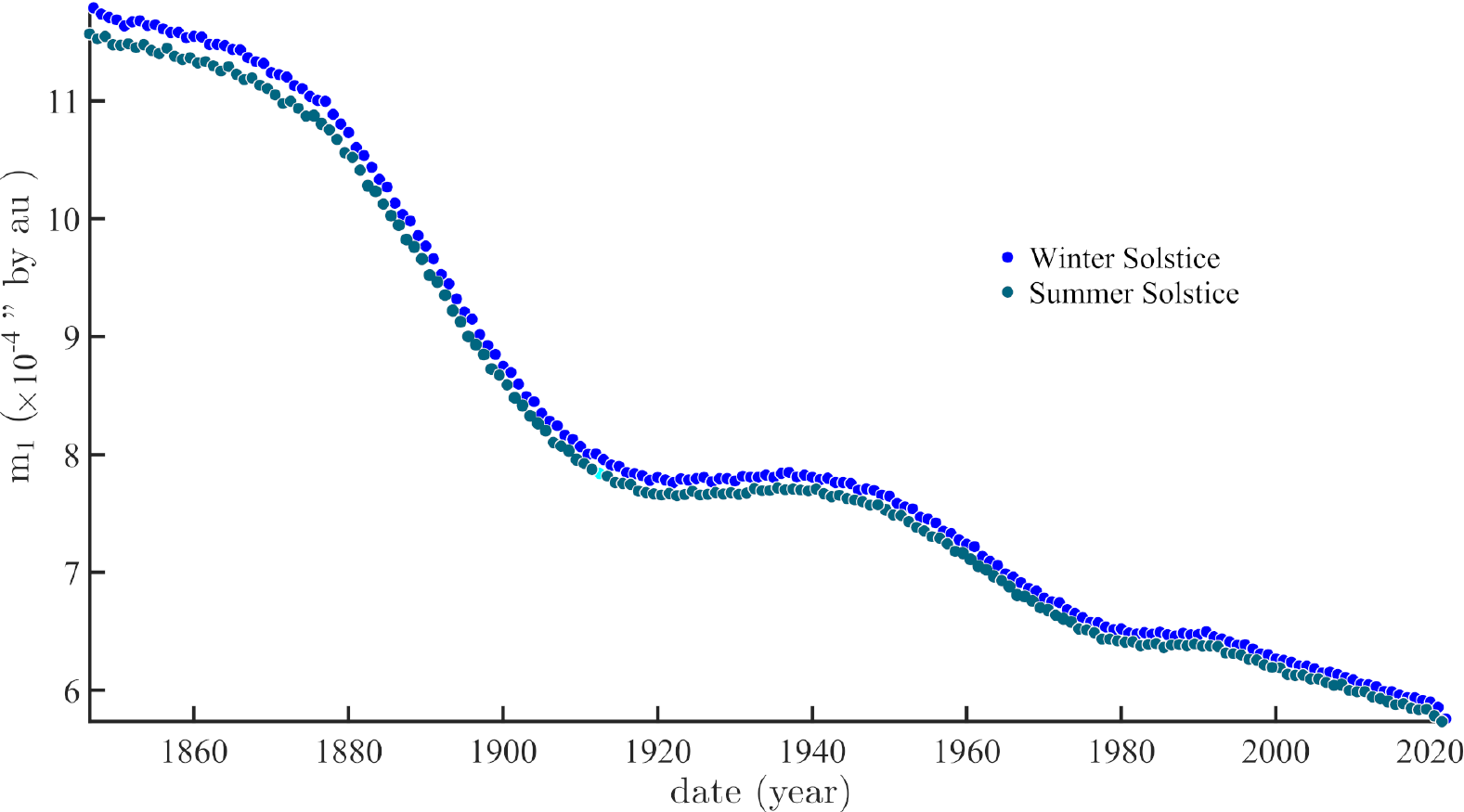}} 	
		\subcaption{Time evolution of Winter and Summer solstices for component $m_1^*$.}
		\label{Fig:07a}
	\end{subfigure}
	\begin{subfigure}[b]{\columnwidth}
		\centerline{\includegraphics[width=\columnwidth]{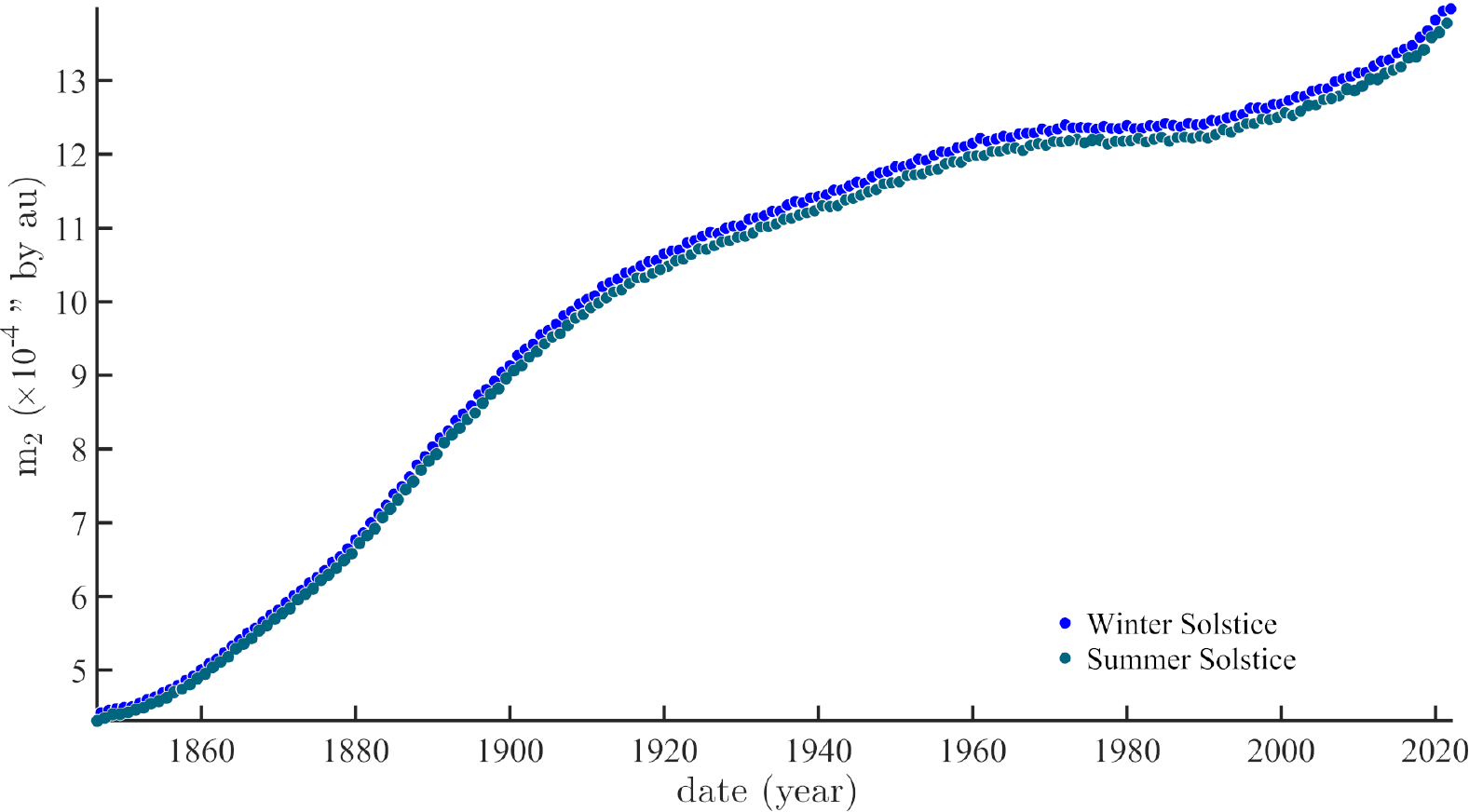}} 	
		\subcaption{Time evolution of Winter and Summer solstices for component $m_2^*$..}
		\label{Fig:07b}
	\end{subfigure}
	\caption{Time evolution of Winter and Summer solstices}
	 \label{Fig:07}
\end{figure}

    We display the drift of the four reference points (solstices and equinoxes) in the ($m_1^*$, $m_2^*$.) plane in Figure \ref{Fig:06b}. The time evolution of the separate coordinates $m_1^*$ and $m_2^*$. for the solstices are shown in Figure \ref{Fig:07a} and \ref{Fig:07b}. \\

\newpage

\newpage
    
\section{Discussion}
    \cite{Milankovic1920} knew that eccentricity, precession and obliquity evolve slowly in time leading to a transition from a warmer to a cooler climate every 11 kyr (the period of precession of equinoxes being the shortest). Figures \ref{Fig:05} to \ref{Fig:07} show that one cannot consider that either the Sun-Earth distance or the hour angle, or the declination, or the daily variation are constant, contrary to what was done by \cite{Milankovic1920}. Both polar motions and length of day are affected by their position on the elliptical orbit, on rather short periods on the order of less than 10 yr up to 1 century or more. The data shown in the present paper demonstrate that parameters in equation (\ref{Eq:01}) evolve in time over these shorter time scales. They may therefore imply some forcing of climate on these same time scales.\\
    
    Figure \ref{Fig:01a} illustrates the differences between five data sets that were supposed when they were compiled to represent the same physical (or quasi-physical) datum, that is mean global surface temperature anomaly. Figure 08a displays the iSSA trends (component 1) of the five HadCrut temperature data sets introduced in section 2-1. The median of the five curves is shown as an inset in Figure \ref{Fig:08a}. Figures \ref{Fig:08b} and \ref{Fig:08c} represent the two other major iSSA components of global mean temperature anomaly, the annual and 60 yr oscillations. The frequencies of the five series are consistent except for the modulated amplitudes of the annual components (Figure  \ref{Fig:08b}), which is puzzling. \\
    
    In Figure  \ref{Fig:09}, we superimpose the time evolution of the Winter (blue) and Summer (green) solstices for component $m_1^*$ from Figure \ref{Fig:07a} and the first iSSA component (trend) of the median of the five HadCrut Center temperature series (red) from Figure \ref{Fig:08a} (note: We have simplified the discussion by using only the m1 polar motion component, the one most clearly linked to the pole’s inclination). With some familiarity with global temperature curves, one can recognize some common features with the evolution of the solstices \cite{LeMouel2020a}. Based on the annual oscillations of the full polar motion ($m_1$, $m_2$ and \textbf{lod}), we have seen that the celestial positions of solstices moved significantly in the past 180 years. To our knowledge, this is the first time one evidences in the observational data what \cite{dAlembert1749} called the apparent precession of solstices (that is seen from Earth) and that \colorREF{Milankovic (1920}, part 2, chapter 2) worked on but on much longer time periods of millions of yearsWe recalled in the introduction the basic equation from \colorREF{Milankovic’s} thesis (1920) that links time variations of heat received at a given location on Earth to solar insolation, known functions of the location coordinates, solar declination and hour angle, with an inverse square dependence on the Sun-Earth distance. We can let W play the role of the heat/energy term in equation (\ref{Eq:01}). The goal is to translate the drift of solstices as a function of distance to the Sun into the geometrical insolation theory of \cite{Milankovic1920}.
    
\section{Conclusion}
    The Earth’s revolution is modified by changes in inclination of its rotation axis, principally due to the actions of the Moon and Sun. Despite the fact that the gravity field is central, the Earth’s trajectory is not closed and the equinoxes drift (by a little less than one minute of arc per year, that is a precession period of some 26 kyr - \cite{dAlembert1749}). For d’Alembert, Lagrange, Laplace and Poincaré who based their reasoning on the action of torques, as in a top, changes in polar motion and revolution are coupled (through the Liouville-Euler system of equations). Re-organization of Earth’s fluid envelopes follows. For Newton, who gives a central role to the inertia tensor, planetary bodies attract (or repel) the oceans (and atmospheres) and this re-organization of masses modifies the rotation axis.\\
    
\colorREF{Milanković (1920}, p.221) argued that the shortest precession period of perihelion (for us solstices) is 20,700 Julian years and that the consequence of this precession is that the Summer solstice in one hemisphere takes place alternately every 11 kyr at perihelion (a warmer Summer) and at aphelion (a cooler Summer). The difference in insolation between maximum and minimum is a function of eccentricity. Milanković assumed that the planetary distances to the Sun and the solar ephemerids are constant. There are now observations that allow one to drop the hypothesis that Sun-Earth distance, the Sun’s declination and hourly angle (equation \ref{Eq:01}) are constant. Thus, we can evaluate whether changes in the position of the rotation axis affect, for instance, atmospheric temperature, that is the main parameter in \colorREF{Milanković’s (1920)} theory of climate. \\
\begin{figure}[H]
    \centering
	\begin{subfigure}[b]{\columnwidth}
		\centerline{\includegraphics[width=\columnwidth]{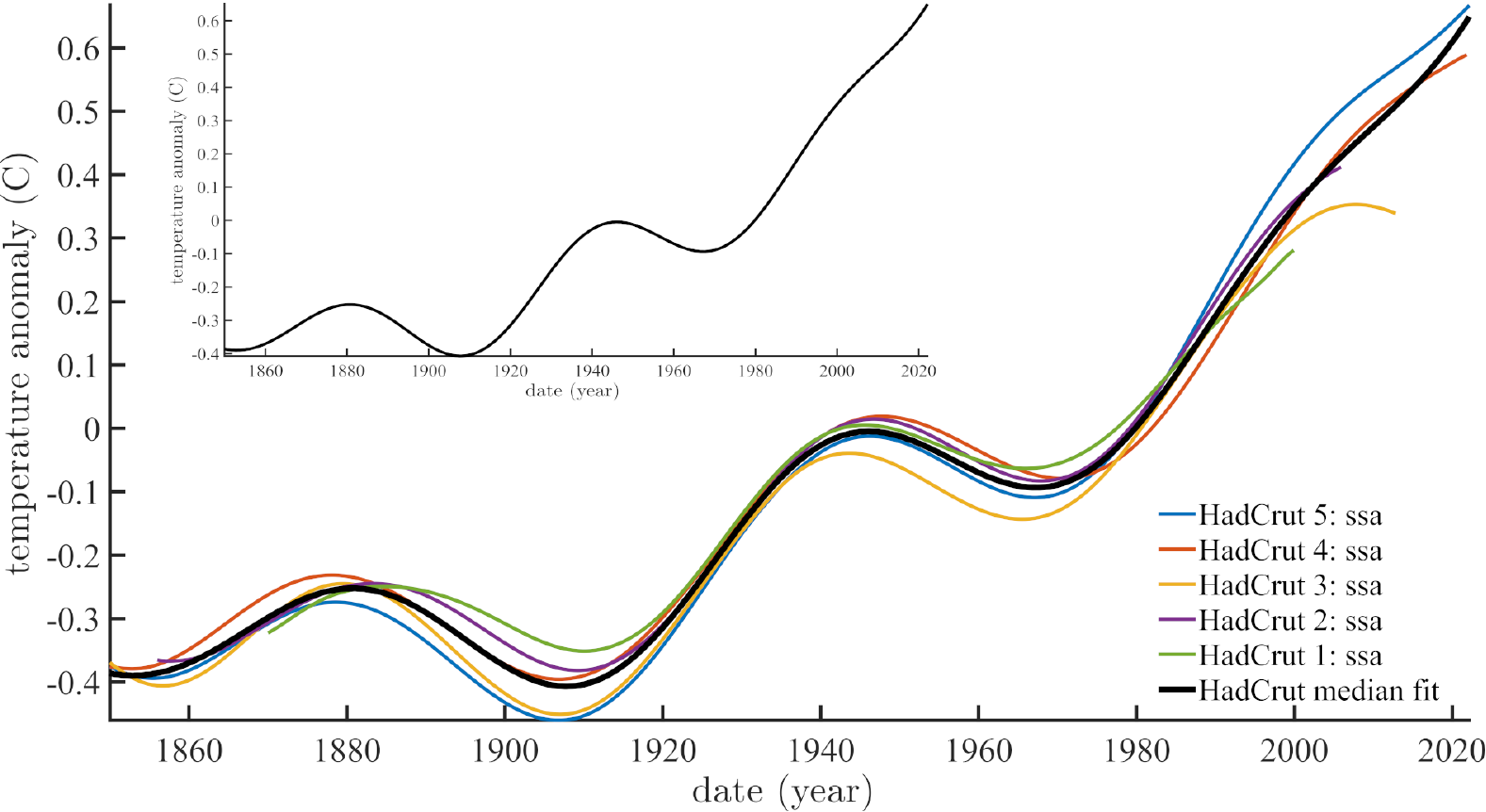}} 	
		\subcaption{The first \textbf{iSSA} components (trends) of the five \textbf{HadCrut} Center temperature series introduced in section 2-1, and their median shown in black among the 5 trends and alone above as an inset.}
		\label{Fig:08a}
	\end{subfigure}
	\begin{subfigure}[b]{\columnwidth}
		\centerline{\includegraphics[width=\columnwidth]{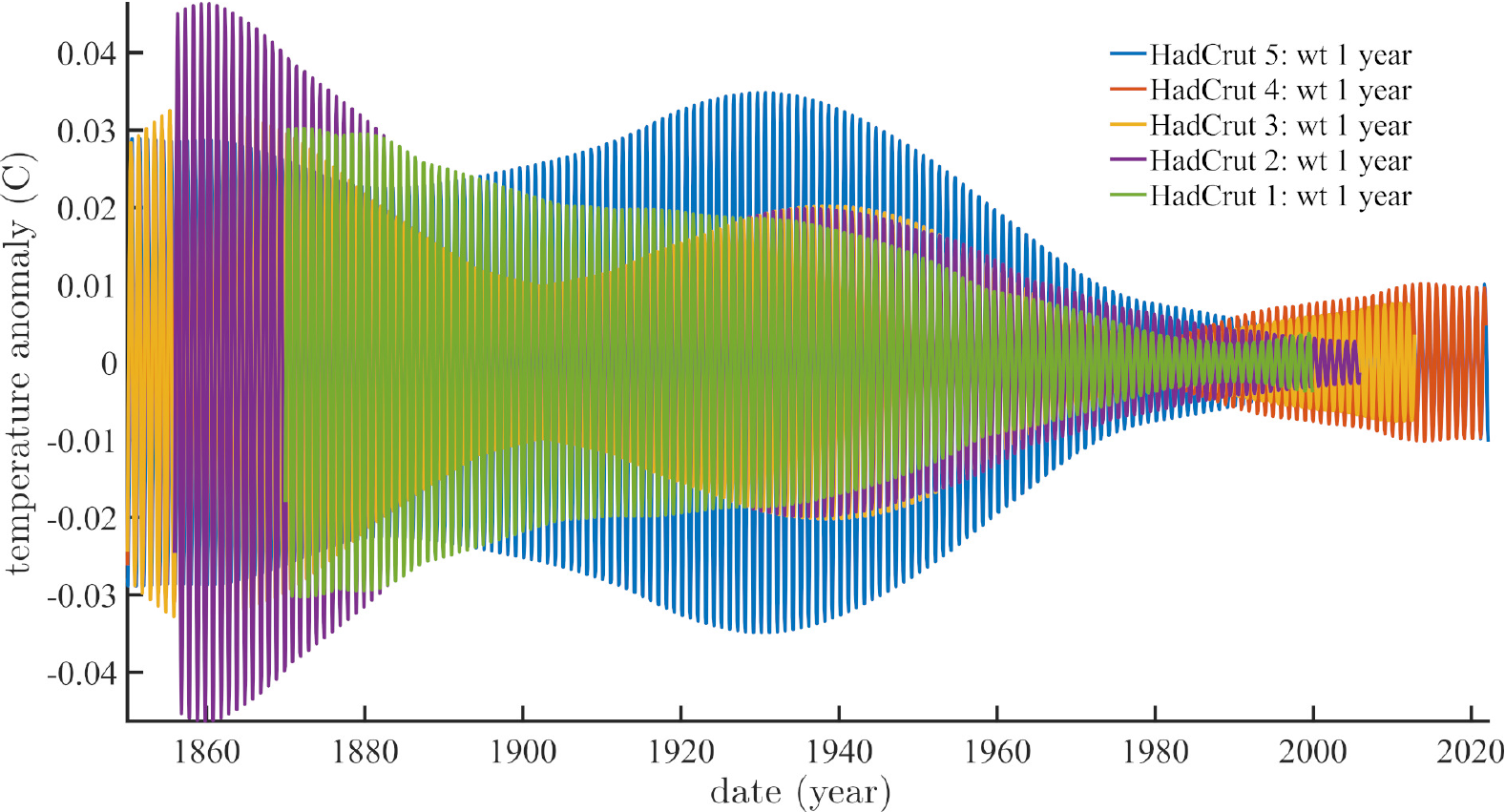}} 	
		\subcaption{Annual iSSA components of the same series as in Figure \ref{Fig:08a}.}
		\label{Fig:08b}
	\end{subfigure}
	\begin{subfigure}[b]{\columnwidth}
		\centerline{\includegraphics[width=\columnwidth]{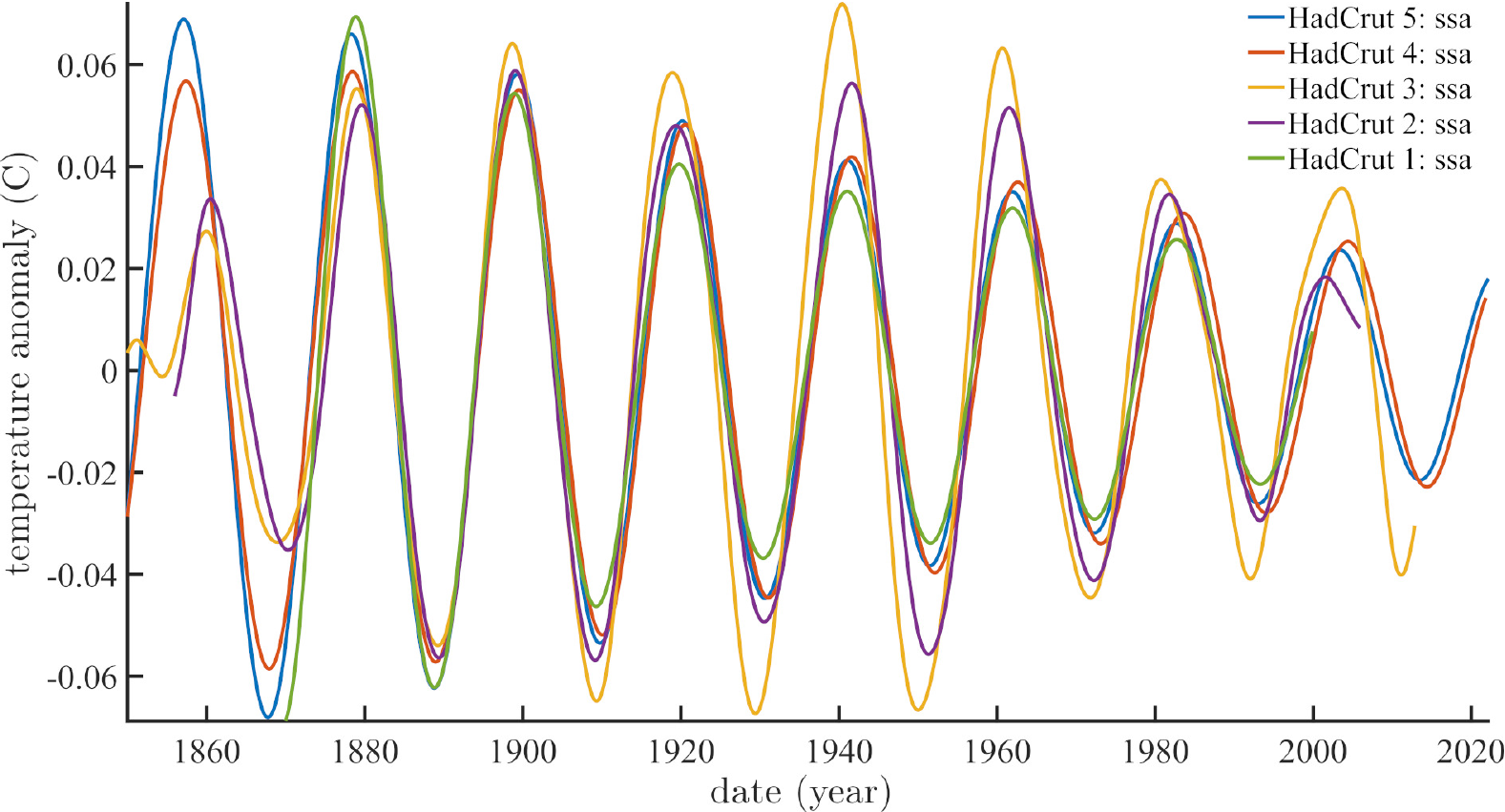}} 	
		\subcaption{Sixty year \textbf{iSSA} components of the same series as in Figure \ref{Fig:08a}.}
		\label{Fig:08c}
	\end{subfigure}
	\caption{Trend, one year and sixty years components extracted from \textbf{HadCrut} curves.}
	 \label{Fig:08}
\end{figure}    

    Both the apparent drift of solstices of Earth around the Sun and the global mean temperature exhibit a strong 60yr oscillation. In the present paper, we have confirmed the finding of a strong iSSA component with 60yr period in global temperatures and in the drift of solstices (Figure \ref{Fig:08c} and \ref{Fig:10}), hence the rotation axis, that has already been encountered in \cite{Lau1995,Chen2004,Groot2011,Sello2011,Zheng2011,Chambers2012,Mazzarella2012,Scafetta2012,Parker2013,dAleo2016,Gervais2016,Veretenenko2019, Scafetta2020,Pan2021}.

\begin{figure}[H]
		\centerline{\includegraphics[width=\columnwidth]{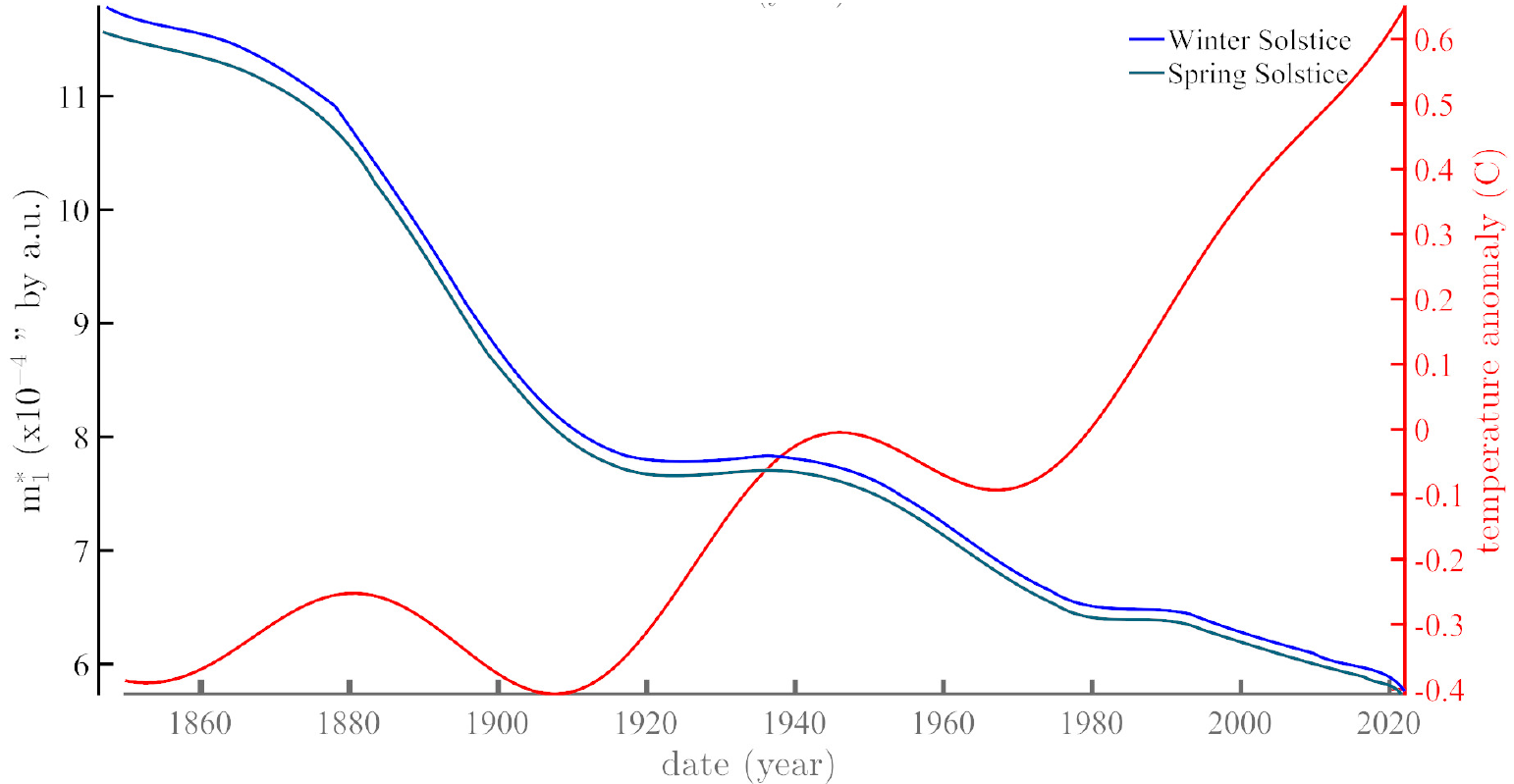}} 	
		\caption{Superposition of the time evolution of the Winter (blue curve) and Summer (green curve) solstices for component $m_1^*$ from Figure \ref{Fig:07a} and the first \textbf{iSSA} component (trend) of the median of the five \textbf{HadCrut} Center temperature series from Figure \ref{Fig:08a} (red curve).}
		\label{Fig:09}
\end{figure}    

    In pursuit of this goal, we obtained the Sun’s ephemerids from 1846 to the present from Institut de Mécanique Céleste et du Calcul des Ephémérides (data set EOP C01 IAU1980). The motions of the rotation pole and variations in rotation velocity were taken from the International Earth Rotation Reference Systems Service. They consist in the couple of coordinates ($m_1$, $m_2$) of motion of the rotation pole PM and the length of day lod. We used the semi-annual lod data provided by \cite{Stephenson1984} for the period 1832-1997, combined with the IERS data, resulting in a mean curve between 1832 and 2022 (Figure \ref{Fig:02c}). For the mean global temperatures, we used five HadCrut data series from the Hadley Center for Climate Prediction and Research in order to estimate the consistence of the data, There are rather significant differences between the five data series, resulting in differences in the dominant spectral peak that shifts from 60 to 80 yr.\\

We have submitted the time series for the rotation, temperature and ephemeris to iterative Singular Spectrum Analysis (iSSA), a method which we have used extensively in a number of recent studies (\cite{LeMouel2019a,LeMouel2019b,LeMouel2020a}). We have checked that the results obtained with iSSA match those obtained with the more common method of continuous wavelets that is widely used in the literature.\\

The main points in space that can be used to monitor the precession of an elliptical orbit are the solstices and equinoxes. The legal and astronomical calendars not being exactly the same, this entails a very small error on their positions (relative variations of their positions between 5 and 9 10-4 ) between 1844 and 2022. \\

Figure \ref{Fig:05a} shows the evolution of the locations of the equinoxes and solstices trajectories as a function of Sun-Earth distance (we call these “Lissajous diagrams”). The closer the Earth is to the Sun, the more the rotation axis straightens; the farther it is, the larger the amplitudes of motions and the flatter the rotation axis. The “fixed dates” of equinoxes and solstices actually drift as a function of time. In order to emphasize the relative amplitudes of the drift and the butterfly-like shape of the Lissajous diagram, we have actually multiplied (m1, m2) by the centered value $d_{SE}^* = d_{SE} – mean(d_{SE})$, yielding ($m_1^*$, $m_2^*$). This also makes it clear that the drift of solstices is larger than that of equinoxes. Polar motion reaches a minimum at the equinoxes when solar attraction is the weakest, and a maximum at the solstices when the Sun, Earth and focus of the ellipse are aligned. \\

Both polar motions and length of day are affected by their position on the elliptical orbit, on rather short periods on the order of less than 10 yr up to 1 to a few centuries. Parameters in equation (\ref{Eq:01}) evolve over these time scales. Some forcing of climate on these same time scales may be expected.
\begin{figure}[H]
		\centerline{\includegraphics[width=\columnwidth]{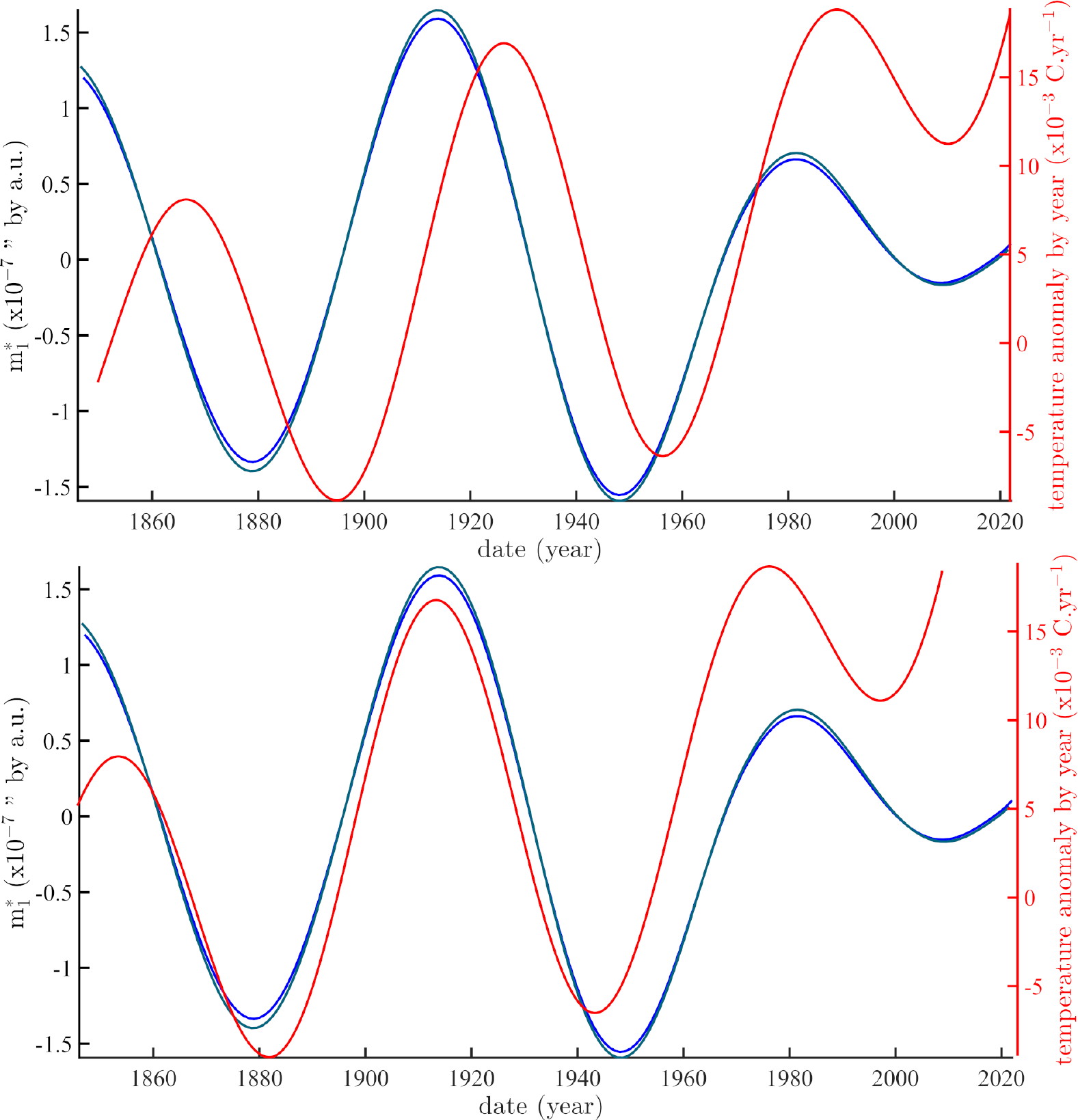}} 	
		\caption{(top) in red the derivative of the \textbf{iSSA} trend of temperature; in green and blue the  inverse square of the drift of solstices. (bottom) a phase quadrature has been applied to the solstices curves above, that is a backward translation of 15 yr (= 60 yr/4).}
		\label{Fig:10}
\end{figure}   

    Despite differences between the five HadCrut temperature data sets, the median of the first iSSA component (trend) appears to be representative (Figure \ref{Fig:08a}). The two other major iSSA components of the five global mean temperature anomaly series, the annual and 60 yr oscillations, are consistent; the modulated amplitudes of the annual components do not agree as well (Figure \ref{Fig:08b}).\\
    
    We have superimposed the time evolution of the Winter and Summer solstices for component $m_1^*$ and the first iSSA component (trend) of the median of the five HadCrut Center temperature series (Figure \ref{Fig:09}). Based on the annual oscillations of the full polar motion (m1, m2 and lod), we have shown that the celestial positions of solstices moved significantly in the past 180 years. To our knowledge, this is the first time one evidences in the observational data what d’Alembert (1749) called the apparent precession of solstices (that is seen from Earth) and that \colorREF{Milankovic (1920}, part 2, chapter 2) worked on, but on much longer time periods of millions of years.\\
        
    One can recognize in Figure \ref{Fig:09} some common features between the evolution of the solstices and the trend of temperatures. We recalled in the introduction the basic equation from \colorREF{Milankovic’s thesis (1920)} that links time variations of heat W received at a given location on Earth to solar insolation, known functions of the location coordinates, solar declination and hour angle, with an inverse square dependence on the Sun-Earth distance. We translate the drift of solstices as a function of distance to the Sun into the geometrical insolation theory of Milankovic (1920). Both the apparent drift of solstices of Earth around the Sun and the global mean temperature exhibit a strong 60yr oscillation (Figure \ref{Fig:08c} and \ref{Fig:10}).\\
    
    It may seem that we navigate between two pitfalls: remembering that correlation does not imply causality at the risk of discounting a potentially interesting relationship or accepting the causality when it does not exist, at the risk of pursuing a non existent theory. We do acknowledge that one should not jump too fast to conclusions, yet the probability of a chance coincidence in Figure \ref{Fig:10} appears very low. Correlation certainly does not imply causality when there is no accompanying model. But in the case studied in this paper, equation (\ref{Eq:01}) can be considered as a model that is widely accepted. Equation (\ref{Eq:01}) links the time derivative of insolation with the inverse square of the Sun-Earth distance. On Figure 10 shifting the inverse square of the 60yr iSSA drift of solstices by 15 years with respect to the first derivative of the 60yr iSSA trend of temperature, that is exactly a quadrature in time, puts the two curves in quasi-exact agreement. This is a case of agreement between observations and a mathematical formulation. This new finding joins a host of recent results that argue in the same direction \cite{Bank2022}; the hypothesis proposes that a forcing by the giant Jovian planets is exerted on a vast number of solar and terrestrial phenomena, including as shown in this paper global surface temperature. This forcing induces a number of responses from the Earth’s rotation axis, hence on climate at many time scales. This is a sort of extension of the Milankovic theory of climate to a range of periods that are much shorter than the $\sim$20 kyr minimum often associated with this theory.
\
\bibliographystyle{aa}

\end{document}